\RequirePackage{fix-cm}
\documentclass[smallextended]{svjour3}       
\smartqed  
\usepackage{graphicx}

\usepackage{epstopdf}

\usepackage{amsmath}
\usepackage{amsfonts}
\usepackage{relsize}
\usepackage{appendix}
\usepackage{amsthm}
\usepackage{enumerate}
\usepackage{color}
\usepackage{natbib}

\DeclareMathOperator*{\argmin}{arg\,min}

\newtheorem{thm}{Theorem}
\newtheorem{lem}[thm]{Lemma}
\newtheorem{pf*}{Proof}

\journalname{.....}

\begin{document}

\title{Nonparametric latency estimation for mixture cure models}
\subtitle{}


\author{Ana L\'opez-Cheda         \and
        M. Amalia J\'acome   \and  Ricardo Cao
}


\institute{Ana L\'opez-Cheda$^{1}$ \at
              \email{ana.lopez.cheda@udc.es}
           \and
           M. Amalia J\'acome$^{2}$ \at
              \email{majacome@udc.es}
           \and
           Ricardo Cao$^{1}$ \at
              \email{rcao@udc.es} \bigskip \\
$^{1}$ Grupo MODES, INIBIC, CITIC, Departamento de Matem\'{a}ticas, Facultade de Infor\-m\'{a}\-ti\-ca, Universidade da Coru\~{n}a, 15071 A Coru\~{n}a, Spain \\
$^{2}$ Grupo MODES, INIBIC, CITIC, Departamento de Matem\'{a}ticas, Facultade de Ciencias, Universidade da Coru\~{n}a, 15071 A Coru\~{n}a, Spain
}

\date{Received: date / Accepted: date}

\maketitle

\begin{abstract}
A nonparametric latency estimator for mixture cure models is stu\-died in this paper. An i.i.d. representation is obtained, the asymptotic mean squared error of the latency estimator is found, and its asymptotic normality is proven. A bootstrap bandwidth selection method is introduced and its efficiency is evaluated in a simulation study. The proposed methods are applied to a dataset of colorectal cancer patients in the University Hospital of A Coru\~{n}a (CHUAC).
\end{abstract}


\section{Introduction}
\label{sec:1}
In the last two decades there has been a remarkable progress in cancer treatments, which led to longer patient survival and improved their quality of life. Consequently, a spate of statistical research to develop cure models arose. These models are a useful tool to analyze and describe cancer survival data, since they express and predict the prognosis of a patient considering, as a no\-vel\-ty, the real possibility that the subject may never experience the event of inte\-rest. Importantly, cure models should not be used in an indiscriminate way \citep[see][]{Farewell2}. They generally require long-term follow-up and large sample sizes, as well as empirical and biological evidence of a nonsusceptible subpopulation. More specifically, they are used to estimate the probability of cure, also known as \emph{incidence}, and the survival function of the uncured po\-pu\-la\-tion denoted by \emph{latency}.\\
\indent Cure models can be split into two major types: the mixture and the nonmixture models.
Mixture cure models were proposed by \cite{Boag}. They consider the survival function as a mixture of two groups of subjects: the susceptible group and the cured group. An important benefit of the mixture cure model is that it allows covariates to have different influence on patients who will experience the event of interest (e.g. death by the cancer under study) and on those who will not. \\
\indent In the literature, the covariate effect is generally expressed parametrically or semiparametrically \citep[see, among others,][]{Farewell1,Goldman1,Kuk,Maller3,Sposto,Chappell,Taylor,Peng1,Sy,Peng3,Yu}. Recently \cite{Louzada} considered re\-current event data in the pre\-sence of a cure fraction. Very few papers exist that use a nonparametric view to deal with the pro\-blem \citep[see][]{Maller1,Laska,Wang}. In the discussion by Van Keilegom to the paper \cite{Gonzalez-Manteiga}, the problem of goodness-of-fit tests for regression models with cured data is briefly consi\-de\-red. A completely nonparametric approach to the mixture cure model was firstly addressed by \cite{Xu}, proposing a non\-pa\-ra\-me\-tric incidence estimator which works with continuous covariates, and proving its consistency and asymptotic normality. This nonparametric incidence estimator was stu\-died later by \cite{LopezCheda}, who obtained an i.i.d. representation, the asymptotically optimal bandwidth and proposed a \mbox{bootstrap} bandwidth selector. Regarding the latency function, a nonparametric estimator was proposed by \cite{LopezCheda}, but no further properties were studied. The present paper contributes to this lacuna studying the asymptotic properties of that nonparametric latency estimator and pro\-po\-sing a bootstrap bandwidth selector. This enables the mixture cure model with covariates to be addressed in a completely nonparametric way.\\
\indent The rest of the paper is organized as follows. Section \ref{sec:2} introduces the notation and presents the nonparametric mixture cure model and the non\-pa\-ra\-me\-tric latency estimation. The asymptotic results for this estimator as well as the required assumptions are also introduced in Section \ref{sec:2}. An i.i.d. re\-pre\-sen\-ta\-tion is presented, an asymptotic expression for the mean squared error is found and the asymptotic normality is established for the nonparametric latency estimator. The problem of choosing the smoothing parameter is addressed in Section \ref{sec:3}, where a bootstrap bandwidth selector is presented. The practical performance of this bootstrap bandwidth selector is assessed by a simulation study in Section \ref{sec:4}. The application of these methods to a colorectal cancer data set is considered in Section \ref{sec:5}. A final Appendix contains the proofs of the theoretical results stated in Section \ref{sec:2}.

\section{Main results}
\label{sec:2}
\subsection{Notation and nonparametric estimators}
To distinguish between cured and uncured subjects we use a binary
indicator: $\nu$. If the subject belongs to the susceptible group we
set $\nu =0$. This means that the individual will experience the
event of interest if followed during enough time. If the subject is
cured we set $\nu =1$. In such a case the event will never be
experienced by that subject. The probability of being cured and the
survival function in the group of uncured patients may depend on a
vector of covariates, $\mathbf{X}$, measured on the subject. Let us
consider $p(\mathbf{x}) = P(\nu =0 | \mathbf{X}=\mathbf{x})$, the
conditional probability of not being cured, and let $Y$ be the time
to the event of interest. If $\nu=1$, we set $Y =\infty$.

We define $F(t|\mathbf{x})=P(Y\leq t |\mathbf{X}=\mathbf{x})$, the conditional distribution function of $Y$. When the cure probability is positive, then the corresponding survival function, $S(t|\mathbf{x})$, is improper. In other terms, $\lim_{t \rightarrow \infty}
S(t|\mathbf{x}) = 1-p(\mathbf{x}) >0$.

Using the conditional survival function for susceptible subjects, $S_0(t|\mathbf{x})=P(Y > t | \mathbf{X}=\mathbf{x}, \nu =0)$, the mixture cure model can be written as:
\begin{equation}  \label{supervivencia_mixtura}
S(t|\mathbf{x})=1 - p(\mathbf{x}) + p(\mathbf{x}) S_0 (t|\mathbf{x}).
\end{equation}
The function $1 - p(\mathbf{x})$ is called the incidence and $S_0(t|\mathbf{x})$ is the latency.

Random right censoring is assumed. The censoring time is denoted by $C$ and $G$ denotes its distribution function ($\bar G$ is its survival function). The va\-ria\-ble $C$ is assumed to be independent of $Y$ given the covariates $\mathbf{X}$. The observed time is defined as $T=\min \{Y,C \}$ and $\delta =1 \{Y \leq C\}$ is the uncensoring indicator. We denote by $H$ the distribution function of $T$. It is clear that $\delta=0$ for
all the cured patients, and also for uncured patients with censored lifetime ($T=C$). From now on we restrict ourselves to the case where $\mathbf{X}$ is a univariate continuous covariate $X$ with density function $m(x)$. As a consequence of the previous definitions and assumptions, the sample is denoted by $\{(X_i, T_i, \delta_i), i=1, \dots, n \}$, which collects i.i.d. observations of the random vector $(X, T, \delta)$. Whenever is needed $(X_{(i)},T_{(i)},\delta_{(i)})$ will denote the observation corresponding to the $i$-th order statistic with respect to the sample $(T_1, T_2, \ldots, T_n)$, where $X_{(i)}$ and $\delta_{(i)}$ are the concommitants of the $X$ and $\delta$-samples.

The conditional distribution, survival and subdistribution functions are denoted by $G(t|x)=P\left( C \leq t|X=x \right)$, $\bar G(t|x)= 1-G(t|x)$, $H(t|x)=P\left( T \leq t | \right . $ $ \left . X=x \right)$, $H^1(t|x)=P\left( T \leq t, \delta =1|X=x \right)$ and  $H_{c,1}(t)=P\left( T < t|\delta = \right. $ $\left. 1 \right)$.

We will consider the nonparametric approach in mixture cure models by \cite{LopezCheda}. It departs from the generalized Kaplan-Meier estimator by 
Beran (1981) to estimate the conditional survival function:
\begin{equation}  \label{S_est}
\hat S_{h}(t|x) = \prod_{T_{(i)} \leq t} \left ( 1 - \frac{\delta_{(i)} B_{h(i)}
(x) }{\sum_{r=i}^n B_{h(r)}(x)}\right ),
\end{equation}
where $B_{h(i)}(x) = K_h (x - X_{(i)}) /    \sum_{j=1}^n K_h(x - X_{(j)}) $ are the Nadaraya-Watson (NW) weights and $K_h(\cdot)=\frac{1}{h}K \left ( \frac{ \cdot}{h} \right )$ the rescaled kernel with bandwidth $h>0$. We denote by $\hat F_h(t|x) = 1 - \hat S_h(t|x)$ the Beran estimator of $F(t|x)$.

Departing from the Beran estimator, \cite{Xu} introduced a kernel type estimator for the incidence function:
\begin{equation}  \label{ec:p_estimation}
1-\hat p_h(x)= \prod_{i=1}^{n} \left( 1 - \frac{\delta_{(i)} B_{h(i)}(x)}{\sum_{r=i}^n B_{h(r)}(x) } \right ) = \hat S_h(T_{\max}^1|x),
\end{equation}
where $T_{\max}^1=\max\limits_{i:\delta_i=1}(T_i)$ is the largest uncensored failure time. These authors proved the consistency and asymptotic normality of $\hat p_h(x)$. \cite{LopezCheda} obtained an i.i.d. representation and the asymptotically optimal bandwidth, proposed a bootstrap bandwidth selector for $\hat p_h(x)$, and introduced the following nonparametric latency estimator:
\begin{equation}  \label{ec:S0_estimation}
\hat S_{0,h} (t|x) = \frac{\hat S_{h}(t|x) - (1 - \hat{p}_h(x))}{\hat{p}_h(x)},
\end{equation}
with $\hat S_{h}(t|x)$, in (\ref{S_est}), the Beran estimator of $S(t|x)$ and $1 - \hat{p}_h(x)$ the estimator by \cite{Xu} in (\ref{ec:p_estimation}). They also addressed identifiability of model (\ref{supervivencia_mixtura}). Note that the optimal bandwidth for $\hat S_{0,h}(t|x)$ is not necessarily the optimal bandwidth for $\hat p_h(x)$. A more general function than (\ref{ec:S0_estimation}) using different bandwidths for the incidence and for the improper survival function:
\begin{equation}  \label{ec:S0_estimation_two_band}
\hat S_{0,h_1,h_2} (t|x) = \frac{\hat S_{h_1}(t|x) - (1 - \hat{p}_{h_2}(x))}{\hat{p}_{h_2}(x)}
\end{equation}
could be considered as an estimator of the latency. However, it does not yield necessarily a proper survival function since its limit as $t$ tends to infinity needs not to be zero. In fact, it is not even guaranteed to be non negative. On the other hand, as it will be shown in Subsection \ref{subsec:two_bandwidths}, the optimal values for $h_1$ and $h_2$ in (\ref{ec:S0_estimation_two_band}) are nearly equal. As a consequence, in this work only the asymptotic properties of the nonparametric latency estimator in (\ref{ec:S0_estimation}), that depends on one unique bandwidth $h$, will be studied. Similar theoretical results, not included in this paper, are easily extended to the estimator in (\ref{ec:S0_estimation_two_band}).

Let us define: $\tau_{S_{0}}(x) = \sup \left\{t:S_{0}(t|x)> 0\right\}$. Since $S( t|x)$ is an improper survival function and $1-H(t|x)=S(t|x) \bar G(t|x)$, then $\tau_{H}(x)=\tau_{G}(x)$, where
$\tau_{H}(x) = \sup \left\{t:H(t|x) < 1\right\}$ and $\tau_{G}(x) = \sup \left\{t:G(t|x) < 1\right\}$.

Let $\tau_{0}=\sup_{x \in D} \tau _{S_{0}}(x) $, where $D$ is the support of $X$. As in \cite{Xu}, we consider
\begin{equation}
\label{ec:cond_tau}
\tau _{0}<\tau _{G}\left( x\right), \forall x \in D.
\end{equation}
The rationale of this condition has been discussed by \cite{LopezCheda}, \cite{Xu} and \cite{Maller1,Maller3}. Note that if the cen\-so\-ring va\-ria\-ble takes values always below a time $\tau_G < \tau_0$, the largest uncensored observation may occur at a time not larger than $\tau_G$ and therefore always before $\tau_0$. \cite{Laska} stated that, for a large sample size, the nonparametric incidence estimator in (\ref{ec:p_estimation}) is an estimator of $1-p(x)+p(x)S_0(\tau_G)$, which is strictly larger than $1-p(x)$. Specifically, as it is mentioned in \cite{Maller1}, consistent estimates of the incidence are possible if and only if there is zero probability of a susceptible individual surviving longer than the largest possible censoring time. That is, condition (\ref{ec:cond_tau}) guarantees that censored subjects beyond the largest observable failure time are cured, since the support of the censoring variable, $C$, is not contained in the support of $Y$, the time to occurrence of the event. Therefore, the nonparametric estimator does not overestimate the true cure rate. A nonparametric test for this condition on the censoring mechanism was proposed by \cite{Maller1} in an unconditional setting, and by \cite{LopezCheda} with covariates.


\subsection{Theoretical results}
\label{subsec:2.2}

The following assumptions are needed to prove the asymptotic results in this
section.

\begin{enumerate}[({A}1)]
\item $X$, $Y$ and $C$ are absolutely continuous random variables.
\item Condition (\ref{ec:cond_tau}) holds.
\item \begin{enumerate}[(a)]
\item Let $I=[x_{1},x_{2}]$ be an interval contained in the support of $m$, and $I_{\delta }=[x_{1}-\delta ,x_{2}+\delta ]$ for some $\delta >0$ such that $0<\gamma =\inf [m\left( x\right) :x\in I_{\delta }]<\sup [m\left( x\right):x\in I_{\delta }]=\Gamma <\infty$ and $0<\delta \Gamma <1$. Then for all $x\in I_{\delta}$ the random variables $Y$ and $C$ are con\-di\-tio\-na\-lly independent given $X=x$.
\item There exist $a,b\in \mathbb{R}$, with $a<b$ satisfying $1 - H(t|x) \geq \theta > 0$ for $(t,x) \in [a,b] \times I_{\delta}$.
\end{enumerate}
\item The first derivative of the function $m(x)$ exists and is continuous in $x\in I_{\delta }$ and the first derivatives with respect to $x$ of the functions $H(t|x)$ and $H^{1}(t|x)$ exist and are continuous and bounded in $\left(t,x\right) \in \lbrack 0,\infty )\times I_{\delta }$.
\item The second derivative of the function $m(x)$ exists and is \mbox{continuous} in $x\in I_{\delta }$ and the second derivatives with respect to $x$ of the functions $H(t|x) $ and $H^{1}(t|x) $ exist and are continuous and bounded in $\left(t,x\right) \in \lbrack 0,\infty )\times I_{\delta }$.
\item The first derivatives with respect to $t$ of the functions $G(t|x)$, $H(t|x)$, $H^{1}(t|x)$ and $S_0(t|x)$ exist and are continuous in $\left( t,x\right) \in \lbrack a,b] \times D$.
\item The second derivatives with respect to $t$ of the functions $H(t|x) $ and $H^{1}(t|x) $ exist and are continuous in $\left(t,x\right) \in \lbrack a,b]\times D$.
\item The second partial derivatives with respect to $t$ and $x$ of the functions $H(t|x)$ and $H^1(t|x)$ exist and are continuous and bounded for $(t,x)\in [0,\infty) \times D$.
\item The first and second derivatives of the distribution and subdistribution functions $H(t)$ and $H_{c,1}(t)$ are bounded away from zero in $[a,b]$. Moreover, $H^{\prime }_{c,1}(\tau_0) > 0$.
\item The functions $H(t|x)$, $S_{0}(t|x)$ and $G(t|x)$ have bounded second-order derivatives with respect to $x$ for any given value of $t$.
\item The kernel function, $K$, is a symmetric density vanishing outside $\left(-1,1\right) $ and the total variation of $K$ is less than some $\lambda <$ $\infty $.
\item The density function of $T$, $f_T$, is bounded away from 0 in $[0,\infty)$.
\end{enumerate}

The proof of Theorem \ref{thm:iid_representation} is based on Theorem 2 in \cite{Iglesias-Perez}, where the assumptions (A1),(A3)-(A9) and (A11)-(A12) are required. Assumptions (A2) and (A10) ensure that Theorem 2 in \cite{Iglesias-Perez}, stated for a fixed $t$ such that $1-H(t|x) \geq \theta > 0 \in [a,b]\times I_\delta$, can be applied to the random value $t=T_{\max}^1$. Assumptions (A4)-(A8) and (A10) are regularity conditions for the functions involved in the proofs and in the asympotic results.

In Theorem \ref{thm:iid_representation} we obtain an i.i.d. representation for $\hat{S}_{0,h}(t|x)$ in (\ref{ec:S0_estimation}).

\begin{thm}
\label{thm:iid_representation}
Suppose that conditions (A1)-(A12) hold, together with $\frac{\ln n}{nh} \rightarrow 0$ and $h=O\left ( \left ( \frac{\ln n}{n}\right )^{\frac{1}{5}} \right )$, then we have an i.i.d. representation for the nonparametric latency estimator for any $t \in [a,b]$:
\begin{equation*}
\hat{S}_{0,h}(t|x)-S_{0}(t|x)=\sum_{i=1}^{n}\eta_{h}(T_{i},\delta
_{i},X_{i},t,x) + O \left ( \left( \frac{\ln n}{nh} \right)^{3/4}\right)
\text{a.s.},
\end{equation*}
\noindent where
\begin{eqnarray*}
\eta_{h}(T_{i},\delta _{i},X_{i},t,x) &=& -\frac{S(t|x)}{p(x)}
\tilde{B}_{h,
i}(x)\xi (T_{i},\delta _{i},t,x)   \\
&-& \frac{(1-p(x))(1-S(t|x))}{p^{2}(x)}\tilde{B}_{h,i}(x)\xi
(T_{i},\delta _{i},\infty ,x),
\end{eqnarray*}
\begin{equation}
\xi \left( T_{i},\delta _{i},t,x\right) =\frac{1\{T_{i}\leq t,\delta _{i}=1\}%
}{1-H(T_{i}|x)} -\int_{0}^{t}\frac{1\{u\leq T_{i}\}dH^{1}(u|x)}{%
\left(1-H(u|x)\right) ^{2}}  \label{ec:xi}
\end{equation}

\noindent and
\begin{eqnarray*}
\tilde{B}_{h,i}(x)=\frac{\frac{1}{nh} K\left( \frac{x-X_{i}}{h}\right) }{m(x)%
}. \label{ec:Btilde}
\end{eqnarray*}
\end{thm}

From Theorem \ref{thm:iid_representation}, important properties of the nonparametric latency estimator can be obtained: the first one is the asymptotic expression of the Mean Squared Error (MSE) given in Theorem \ref{thm:AMSE}, and the second one is the asymptotic normality, shown in Theorem \ref{thm:asymptotic_normality}. But first some notation will be introduced. Let us define
\begin{eqnarray}
\Phi(y,t,x) &=&E\left[ \xi (T,\delta ,t,x)|X=y\right],  \label{Phi} \\
\Phi_{1}(y,t,x) &=&E\left[ \xi ^{2}(T,\delta ,t,x)|X=y\right]   \label{Phi_1}
\end{eqnarray}%
and
\begin{equation*}
\Phi_{2}(y,t,x)=E\left[ \xi (T,\delta ,t,x)\xi (T,\delta ,\infty ,x)|X=y%
\right] ,
\end{equation*}%
\noindent with $\xi $ in (\ref{ec:xi}). The asymptotic bias and variance of the latency estimator will be expressed in terms of the following functions:
\begin{eqnarray}
B_{1}\left( t,x\right)  &=&\frac{S(t|x)}{p(x)m(x)}\left( \Phi ^{\prime
\prime }\left( x,t,x\right) m(x)+2\Phi ^{\prime }\left( x,t,x\right)
m^{\prime }(x)\right) ,  \label{ec:B1} \\
B_{2}\left( t,x\right)  &=&\frac{(1-p(x))(1-S(t|x))}{p^{2}(x)m(x)} \notag \\
&\times& \left( \Phi ^{\prime \prime }\left( x,\infty ,x\right) m(x)+2\Phi ^{\prime }\left(
x,\infty ,x\right) m^{\prime }(x)\right) ,  \label{ec:B2}
\end{eqnarray}%
\noindent where 
\begin{equation*}
\Phi \left( y,t,x\right) =\int_{0}^{t}\frac{dH^{1}\left( v|y\right) }{1-H(v|x)}-\int_{0}^{t}(1-H(v|y))\frac{dH^{1}(v|x)}{\left( 1-H(v|x)\right)^{2}},
\end{equation*}
\noindent and $\Phi ^{\prime }$ and $\Phi ^{\prime \prime }$ are the derivatives of $\Phi (y,t,x)$ with respect to $y$. Furthermore,
\begin{eqnarray}
V_{1}\left( t,x\right)  &=&\left( \frac{S(t|x)}{p(x)}\right) ^{2}\frac{\Phi
_{1}(x,t,x)}{m(x)},  \label{ec:V1} \\
V_{2}\left( t,x\right)  &=&\left( \frac{(1-p(x))(1-S(t|x))}{p^{2}(x)}\right)
^{2}\frac{\Phi _{1}(x,\infty ,x)}{m(x)},  \label{ec:V2} \\
V_{3}\left( t,x\right)  &=&\frac{(1-p(x))S(t|x)(1-S(t|x))}{p^{3}(x)m\left(
x\right) }\Phi _{2}(x,t,x)  \label{ec:V3}
\end{eqnarray}%
\noindent respectively, where
\begin{equation*}
\Phi _{1}(x,t,x)=\Phi _{2}(x,t,x)=\int_{0}^{t}\frac{dH^{1}\left( v|x\right)
}{\left( 1-H(v|x)\right) ^{2}}.
\end{equation*}%

Note that, except for some constants, $B_1(t,x)$ in (\ref{ec:B1})
and $B_2(t,x)$ in (\ref{ec:B2}) are the dominant terms of the asymptotic bias
of the estimators $\hat S_h$ and $1-\hat p_h$ in (\ref{S_est}) and (\ref{ec:p_estimation}), respectively. Similarly, the terms $V_1(t,x)$ in (\ref{ec:V1}) and $V_2(t,x)$ in (\ref{ec:V2}) are the dominant terms of the corresponding asymptotic variances of $\hat S_h$ and $1-\hat p_h$. Finally, $V_3(t,x)$ in (\ref{ec:V3}) accounts for the
covariance of both estimators.

\begin{thm}
\label{thm:AMSE}
Under assumptions (A1)-(A10), if $\frac{\ln n}{nh} \rightarrow 0$ and $h=O\left ( \left ( \frac{\ln n}{n}\right )^{\frac{1}{5}} \right )$, then the asymptotic mean squared error of the latency estimator is
\begin{equation}
AMSE(\hat{S}_{0,h}(t|x))=\frac{h^{4}}{4}d_{K}B^{2}\left( t,x\right) +\frac{%
c_{K}}{nh}V\left( t,x\right) +o(h^{4})+O\left( \frac{1}{n}\right),\label{th2:amse}
\end{equation}%
\noindent where $d_{K}=\int v^{2}K(v)dv$, $c_{K}=\int K^{2}(v)dv$,
\begin{eqnarray}
B\left( t,x\right) &=&B_{1}\left( t,x\right) +B_{2}\left( t,x\right) ,
\label{ec:B} \\
V\left( t,x\right) &=&V_{1}\left( t,x\right) +V_{2}\left( t,x\right)
+2V_{3}\left( t,x\right),  \label{ec:V}
\end{eqnarray}%
\noindent with $t \in [a,b]$, $B_{1}$, $B_{2}$, $V_{1}$, $V_{2}$ and $V_{3}$ in (\ref{ec:B1})-(\ref{ec:V3}).
\end{thm}

\begin{thm}
\label{thm:asymptotic_normality} Under assumptions (A1)-(A10), if $h \rightarrow 0$ and $\frac{(\ln n)^3}{nh} \rightarrow 0$, it follows that, for any $t \in [a,b],$

a) If $nh^{5} \rightarrow 0$, then
\begin{equation*}
\sqrt{nh} \left( \hat{S}_{0,h}(t|x)-S_{0}(t|x)\right) \xrightarrow{d}%
N\left(0,V\left( t,x\right) c_{K} \right).
\end{equation*}%

b) If $n h^5 \rightarrow C^5 > 0$, then
\begin{equation*}
\sqrt{nh}\left( \hat{S}_{0,h}(t|x)-S_{0}(t|x)\right)
\xrightarrow{d}N\left( B\left( t,x\right) C^{5/2} d_{K} ,V\left(
t,x\right) c_{K} \right).
\end{equation*}
\end{thm}

\section{Bandwidth selection}
\label{sec:3}

From Theorem \ref{thm:AMSE}, the asymptotic mean integrated squared
error of the latency estimator is:
\begin{equation*}
AMISE(\hat{S}_{0,h}(\cdot |x))=\frac{1}{4}d_{K}^{2}h^{4}\int B^{2}\left(
t,x\right) dt+\frac{c_{K}}{nh}\int V\left( t,x\right) dt+o(h^{4})+O\left(
\frac{1}{n}\right) ,
\end{equation*}
where $B(t,x)$ and $V(t,x)$ are defined in (\ref{ec:B}) and (\ref{ec:V}).
The bandwidth which minimizes the asymptotic mean integrated squared error
is
\begin{equation*}
h_{AMISE}(x)=\left( \frac{c_{K}\int V(t,x)dt}{d_{K}^{2}\int B^{2}(t,x)dt}%
\right) ^{1/5}n^{-1/5},
\end{equation*}%
which depends on plenty of unknown functions that are very hard
to estimate. Consequently we propose to select the bandwidth using the
bootstrap method.


\subsection{Bootstrap bandwidth selector}
\label{sec:3_1}

The bootstrap bandwidth selector is the minimizer of the bootstrap
version of the mean integrated squared error (MISE), that can be approximated, using Monte Carlo, by:
\begin{equation}
MISE_{x,g}^*(h) \simeq \frac{1}{B} \sum_{j=1}^B \int \left ( \hat
S_{0,h}^{*(j)}(t|x) - \hat S_{0,g}(t|x) \right )^2 w(t) dt,
\label{MISE*MC}
\end{equation}
where $w$ is an appropriate weight function, $\hat
S_{0,h}^{*(j)}(t|x)$ is the kernel estimator of $S_0(t|x)$ in (\ref%
{ec:S0_estimation}) using bandwidth $h$ and based on the $j$-th
bootstrap resample, and $\hat S_{0,g}(t|x)$ is the same estimator
computed with the original sample and pilot bandwidth $g$.

We consider an unconditional censoring bootstrap resampling,
assu\-ming that $G(t|x)=G(t),\;\forall x,t$:

\begin{enumerate}
\item[1.] {For $i= 1, 2, \ldots , n$, generate $C_i^*$ from the product-limit estimator $\hat
G^{KM}$.}

\item[2.] {For $i= 1, 2, \ldots , n$, fix the bootstrap covariates $X_i^*=X_i$
and generate $Y_i^*$ from $\hat S_{0,g}(\cdot |X_i^*)$ with probability $%
\hat p_g(X_i^*)$, and $Y_i^* = \infty$ otherwise.}

\item[3.] {Finally, define $T^*_i=\min \{Y^*_i, C^*_i \}$ and $\delta^*_i =
1\{Y^*_i \leq C^*_i\}$ for $i= 1, 2, \ldots , n$.}

\item[4.] {Repeat Steps 1-3 above $B$ times to generate bootstrap resamples
of the form $\big \{ ( X_1^{(b)}, T_1^{*(b)}, \delta_1^{*(b)}), \dots, ( X_n^{(b)}, T_n^{*(b)}, \delta_n^{*(b)} ) %
\big \}$, $b=1, \dots, B$.}

\item[5.] {For the $b$-th bootstrap resample $(b = 1, 2 \ldots ,B)$, compute
$\hat S_{0,h}^{\ast (b)}(t|x)$ with bandwidth $h_l \in \{h_1,\ldots,h_L\}$%
.}

\item[6.] {With the original sample and pilot bandwidth $g$, compute $\hat
S_{0,g}(t|x)$.}

\item[7.] {For each bandwidth $h_l$ in $\{h_1, \dots, h_L\}$, compute the
Monte Carlo appro\-xi\-ma\-tion of $MISE_{x,g}^*(h_l)$ as in
(\ref{MISE*MC}).}
\item[8.] {Find $h^*_x=\argmin\limits_{h_l \in \{h_1,\ldots,h_L\}}MISE^*_{x,g}(h_l)$.}
\end{enumerate}

\section{Simulation study}

\label{sec:4}

Good practical behavior of the nonparametric latency estimator has been preliminary reported by \cite{LopezCheda}. The purpose of this simulation study is to assess the performance of the bootstrap bandwidth selector for the nonparametric latency estimator. We will work with the same two models considered by \cite{LopezCheda}. For both models, the censoring times are generated according to an exponential distribution with mean $10/3$ and the covariate $X$ has a $U(-20,20)$ distribution.

\paragraph{Model 1}
The probability of not being cured is a logistic function and the latency is close to fulfill the proportional 
hazards model, truncated to guarantee condition (\ref{ec:cond_tau}):
\begin{equation*}
p(x) = \frac{ \exp ( \beta_0 + \beta_1 x ) }{ 1 + \exp (\beta_0 +
\beta_1 x) }
\text{ and }
S_0(t|x) =\dfrac{ \exp(-\lambda(x) t) -
\exp(-\lambda(x) \tau_0) }{1 - \exp(-\lambda(x) \tau_0)}1\{t \leq
\tau_0 \},
\end{equation*}
with $\beta_0 = 0.476$ and $\beta_1 = 0.358$, $\tau_0 = 4.605$ and
$\lambda \left( x\right) =\exp \left( (x+20) /40 \right) $. A
percentage of 54\% of the patients are censored and 47\% are cured.

\paragraph{Model 2}
The probability of not being cured is
\begin{equation*}
p(x) = \frac{\exp \left ( \beta_0 + \beta_1 x + \beta_2 x^2 +
\beta_3 x^3 \right ) }{ 1 + \exp \left (\beta_0 + \beta_1 x +
\beta_2 x^2 + \beta_3 x^3 \right) },
\end{equation*}
with $\beta_0 = 0.0476$, $\beta_1 = - 0.2558 $, $\beta_2 = - 0.0027$
and $\beta_3 = 0.0020$, and $S_0 (t|x) = \frac{1}{2} \left (
\exp(-\alpha(x) t^5 ) + \exp(-100 t^5) \right )$ with $\alpha(x) =
\frac{1}{5} \exp((x+20)/40)$. In this case, the percentages of cure
and censoring are slightly higher than for Model 1: around 62\% of
the individuals are censored and 53\% are cured.

In order to approximate the bootstrap version of the $MISE_x$ of the nonparametric latency
estimator, $m=1000$ trials and $B=200$ bootstrap resamples of sizes
$n=50$, $n=100$ and \mbox{$n=200$} were drawn and used the
Epanechnikov kernel. We considered a grid of $35$ bandwidths (from $5$ to $100$)
equispaced on a logarithmic scale. Note that, although the covariate
$X\in U[-20,20]$, we only work with $x \in [-10,20]$. The reason is that $p(x)\simeq
0$ for $-20\leq x\leq -10$. This implies that almost all the subjects are cured, and therefore the estimation of the survival function of the uncured population can not be obtained. Similarly as for the nonparametric incidence estimator \citep[see][]{LopezCheda}, the effect of the choice of the pilot bandwidth, $g$, on the bootstrap bandwidth, $h^*_x$, is very weak. In this simulation study, we considered the same naive pilot bandwidth selector, $g = C(X_{[n]} - X_{[1]}) \cdot n^{-1/9}$, as in \cite{LopezCheda}, with $C=0.75$, and where $X_{[n]}$ ($X_{[1]}$) is the maximum (minimum) value of the observed values of the covariate $X$.

\begin{figure}
\includegraphics[width=0.5\textwidth]{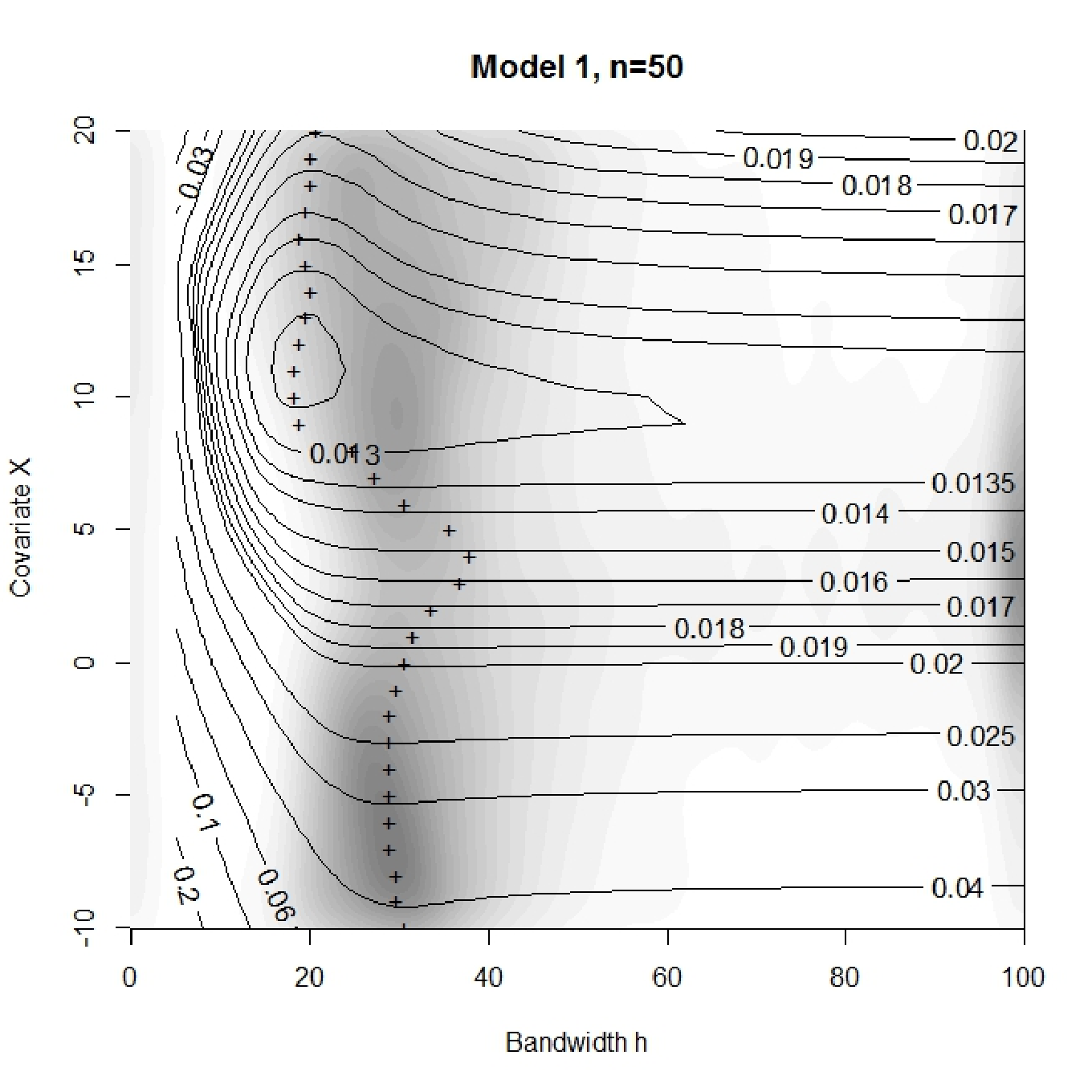}
\includegraphics[width=0.5\textwidth]{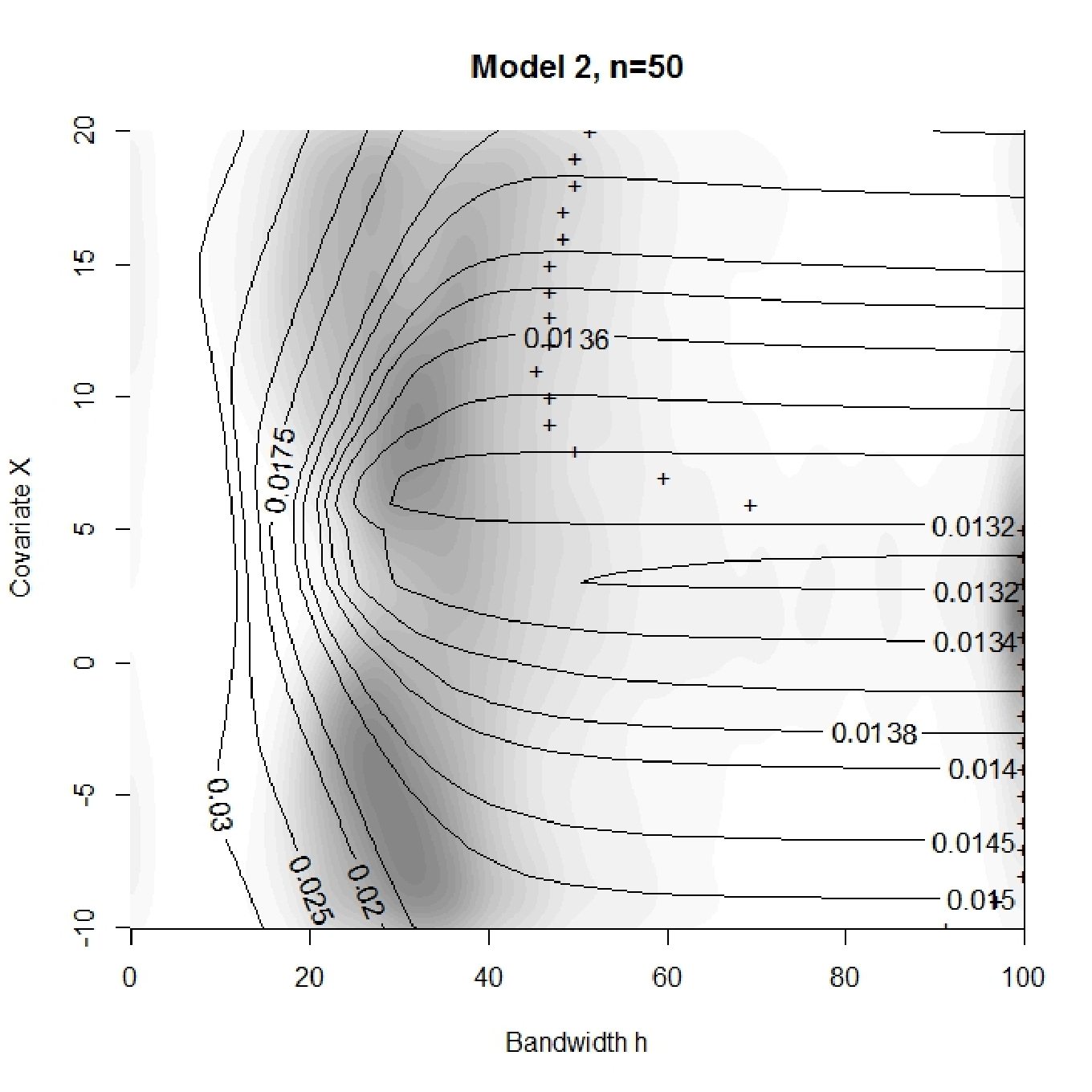} \\
\includegraphics[width=0.5\textwidth]{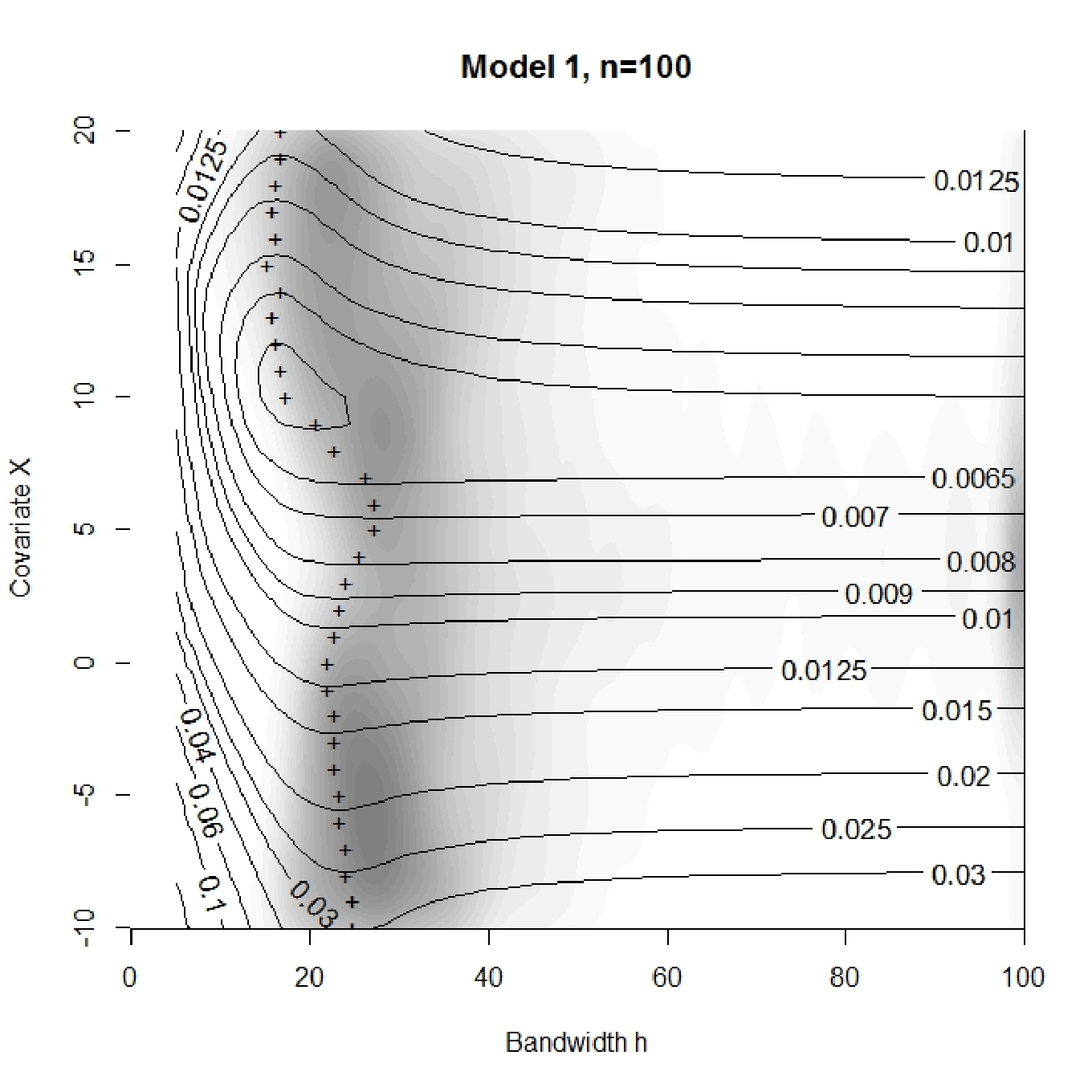}
\includegraphics[width=0.5\textwidth]{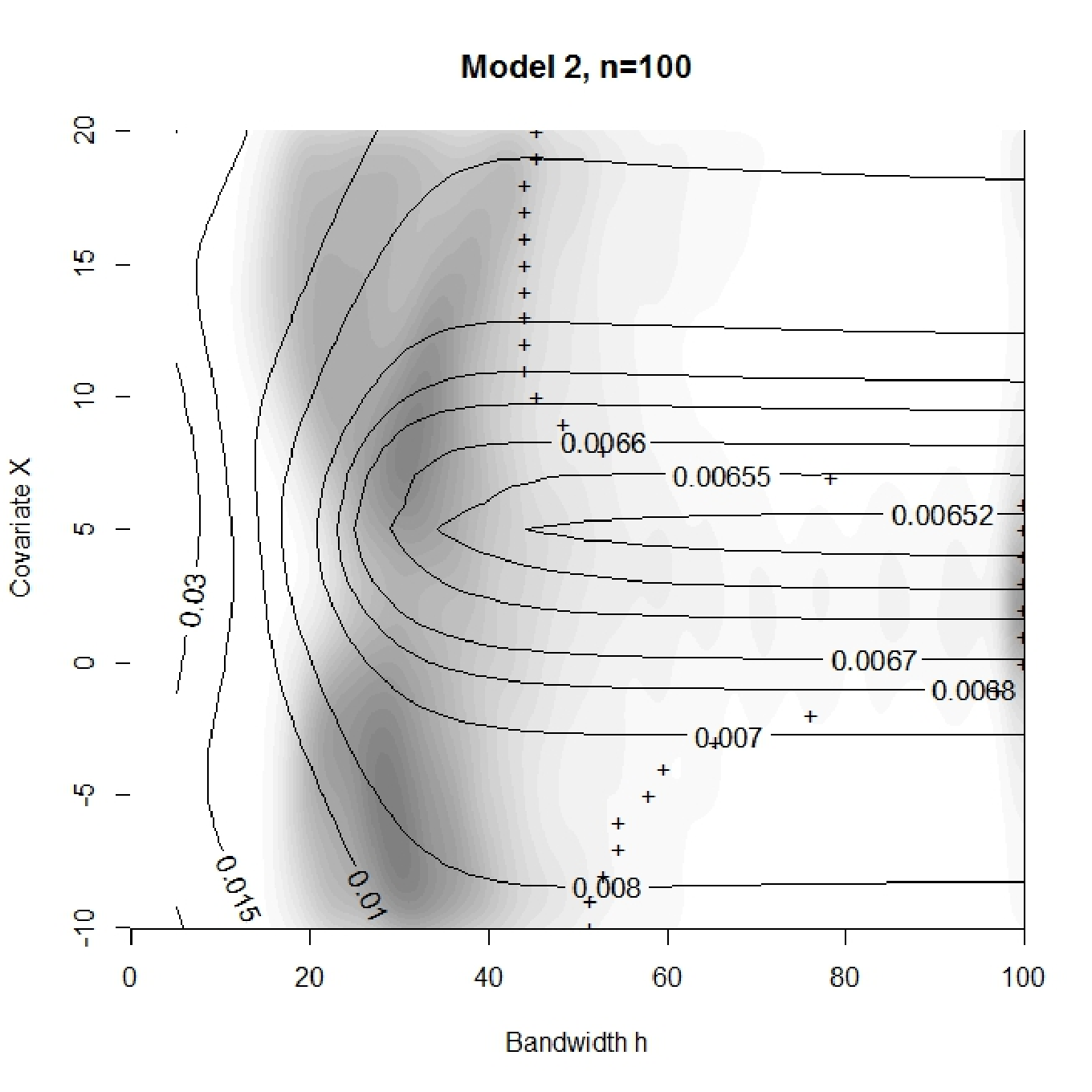} \\
\includegraphics[width=0.5\textwidth]{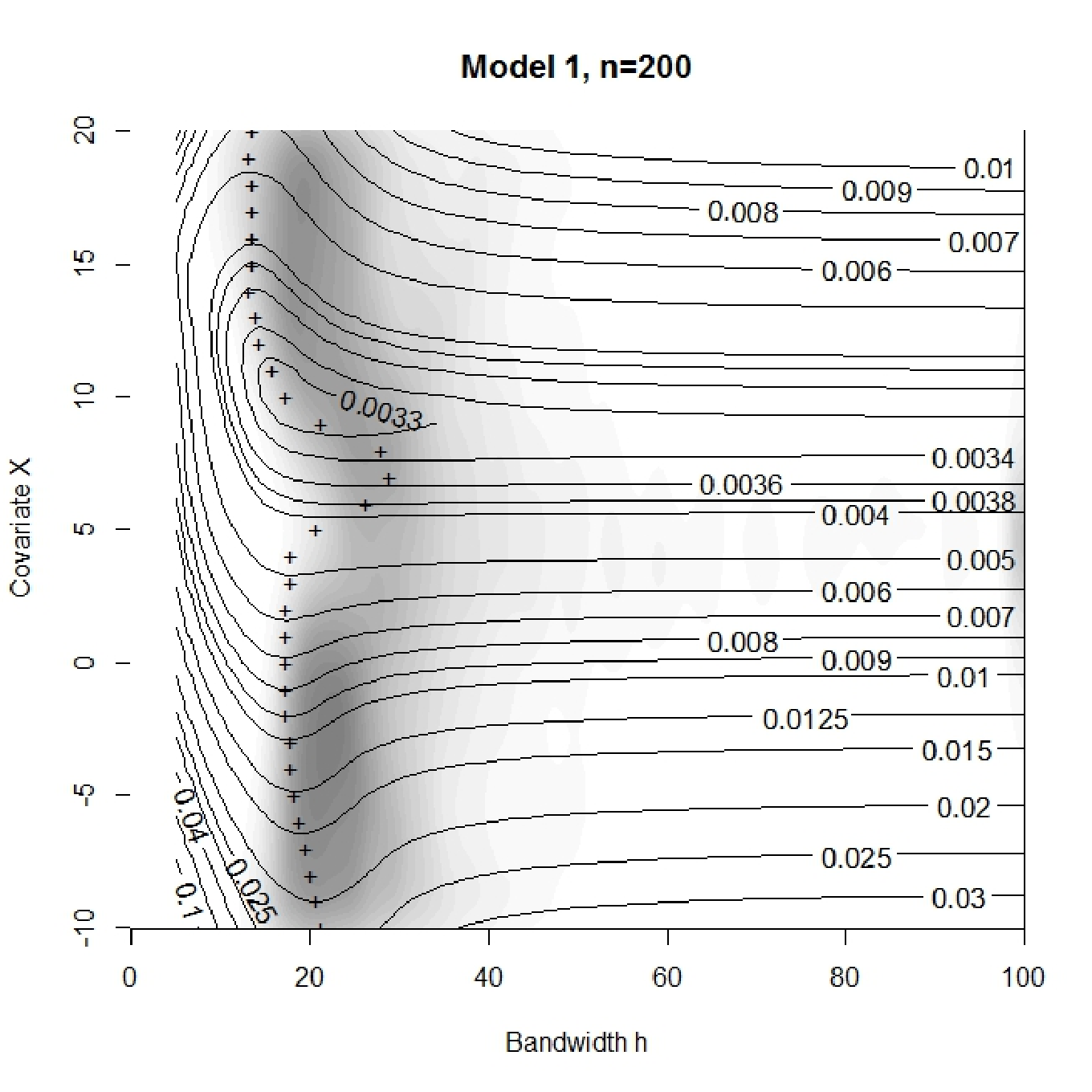}
\includegraphics[width=0.5\textwidth]{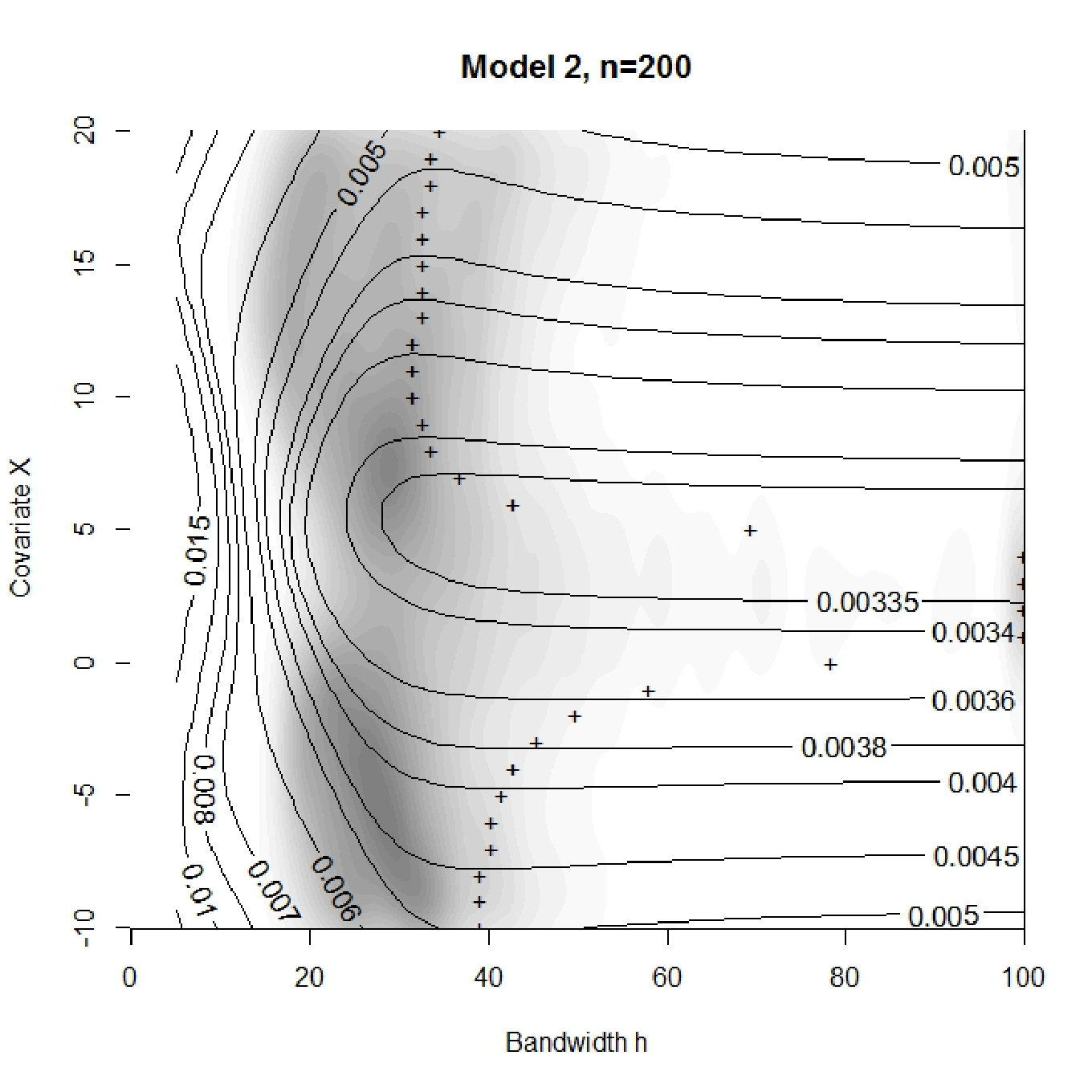} \\
\caption{MISE contour plot depending on the bandwidth and on the covariate, for Model 1 (left) and Model 2 (right), with sample sizes $n=50$ (top), $n=100$ (center) and $n=200$ (bottom). The density of the bootstrap bandwidth is displayed in grayscale and the $h_{MISE}$ bandwidth, for each covariate value, is represented with crosses.}
\label{fig:MISExh}
\end{figure}

In Figure \ref{fig:MISExh} the density of the bootstrap bandwidths, $h^*_x$, is compared with the optimal $h_{MISE,x}$ bandwidth. 
The $MISE$ values obtained con\-si\-de\-ring these bandwidths are also shown. It is noteworthy that $MISE(\hat S_{0,h}(\cdot|x))$, and consequently $MISE^*_{x,g}(h)$, is almost constant in a very wide interval around its minimizer. This feature implies that very different bandwidths could yield very similar good estimates in terms of MISE. We can appreciate how the \mbox{bootstrap} bandwidth might be larger (smaller) than $h_{MISE}$ in Model 1 (Model 2), for most of the covariate values, reflected in a very little difference in terms of MISE between the estimates with the optimal and the bootstrap bandwidths.

\subsection{Results when using two bandwidths to estimate $S_0$}
\label{subsec:two_bandwidths}

\begin{figure}
\includegraphics[width=0.5\textwidth]{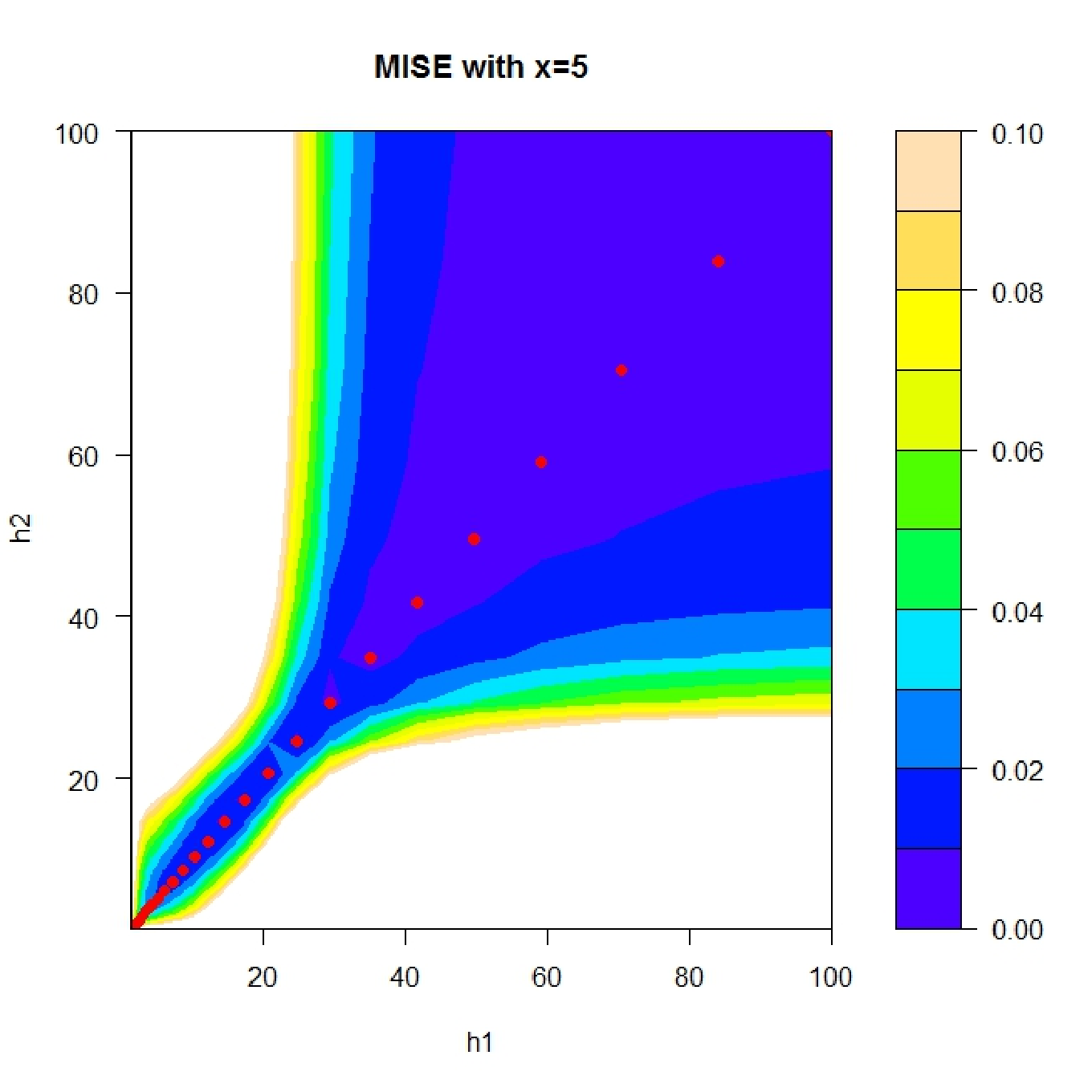} %
\includegraphics[width=0.5\textwidth]{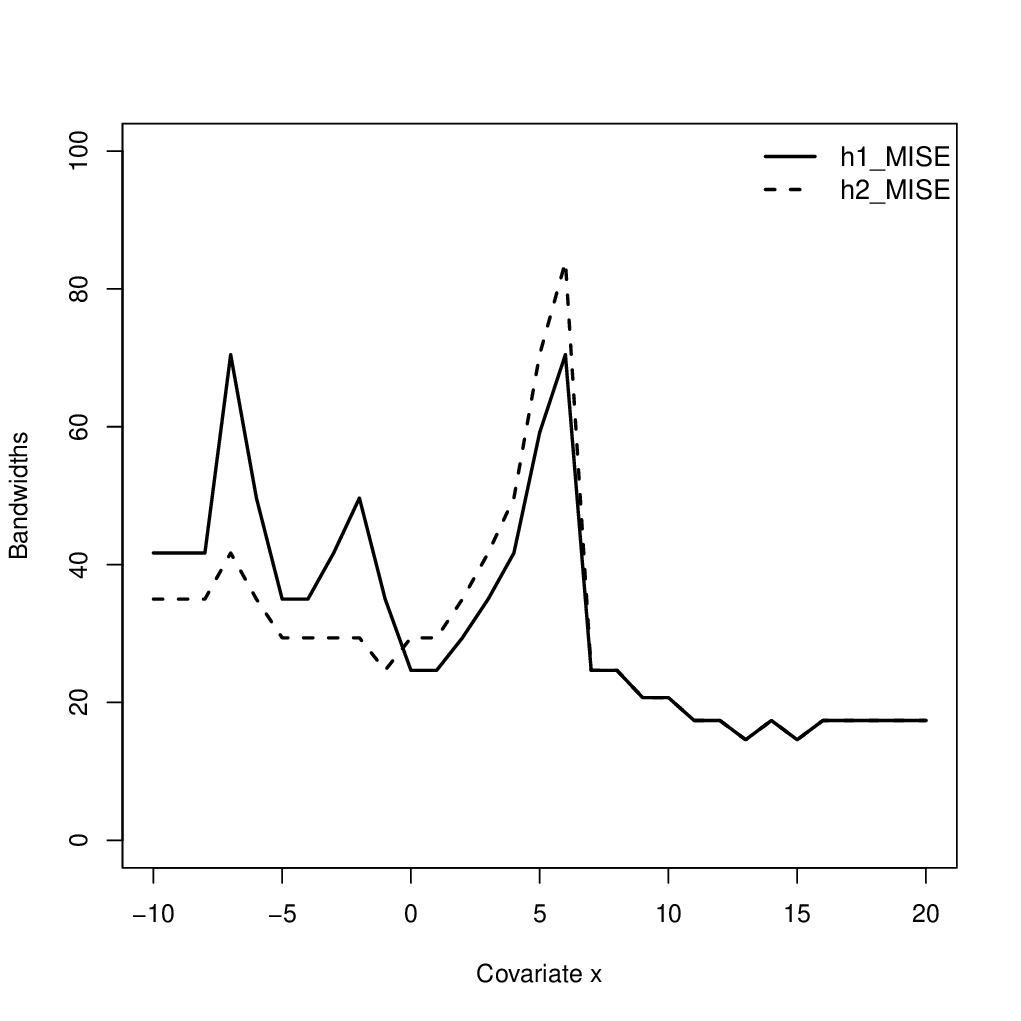}
\caption{$MISE(h_1, h_2)$ of $\hat S_{0,h_1,h_2}$ for $x=5$ and the
grid of bandwidths (equispaced on a logarithmic scale) where
$h_1=h_2$ are represented with red dots (left), and optimal $(h_1,h_2)$
bandwidths, in terms of MISE (right).} \label{fig:h1h2_MISE}
\end{figure}

We will present some results for the latency estimator in (\ref{ec:S0_estimation_two_band}), that is, if two different
bandwidths are considered: $h_1$ for the incidence and $h_2$ for the improper survival function $S$. Note that, for the sake of brevity, we only work with Model 1 and sample size $n=100$, considering $m=1000$ samples. Figure \ref{fig:h1h2_MISE} (left) shows the MISE, approximated by Monte Carlo, of the nonparametric latency estimator $\hat S_{0,h_1,h_2}(t|x)$ in (\ref{ec:S0_estimation_two_band}) as a function of $(h_1,h_2)$ for the covariate value $x=5$ (the MISE for other values of $x$ is similar, not shown). We can see that the minimum MISE (purple color) is reached around the diagonal, that is, when $h_1=h_2$. Figure \ref{fig:h1h2_MISE} (right) provides the optimal bandwidths $(h_1,h_2)$ as a function of $x$. Note that for most of the covariate values both optimal bandwidths are very similar, being even equal for the values of $x$ larger than $5$. \\
Therefore, as pointed out in Section \ref{sec:2}, little efficiency is lost when considering one only bandwidth $h_1 = h_2$ to estimate $S_0$, while this guarantees that the resulting estimator is a proper survival function.

\section{Application to colorectal cancer data}

\label{sec:5} The proposed method was applied to the dataset used in \cite{LopezCheda}, composed of $414$ colorectal cancer
patients from CHUAC (Complejo Hospitalario Universitario de A
Coru\~{n}a), Spain. The variable of interest is the follow-up time,
in months, since the diagnostic until death. Two covariates are
considered: the stage (from $1$ to $4$) and the age (from $23$ to
$103$). The percentage of censoring varies from $30$\% to almost
$71$\%, depending on the stage. In Table 1 we show a summary of the
data set.

\begin{center}
\begin{tabular}{cccc}
Stage & Number of patients & Number of censored data & \% Censoring \\ \hline
1 & 62 & 44 & 70.97 \\
2 & 167 & 92 & 55.09 \\
3 & 133 & 53 & 39.85 \\
4 & 52 & 16 & 30.77 \\ \hline
& 414 & 205 & 49.52\\
\multicolumn{4}{c}{Table 1: Colorectal cancer patients from CHUAC} \\

\end{tabular}
\end{center}

Due to the small sample sizes in each stage, the results are presented in two groups: Stages $1-2$ and Stages $3-4$. Note that $B=200$ bootstrap resamples are drawn. Similarly to the simulation study in Section \ref{sec:4}, we considered a grid of $35$ bandwidths from $h_1=5$ to $h_{35}=100$ equispaced on a logarithmic scale.

\begin{figure}
\includegraphics[width=0.5\textwidth]{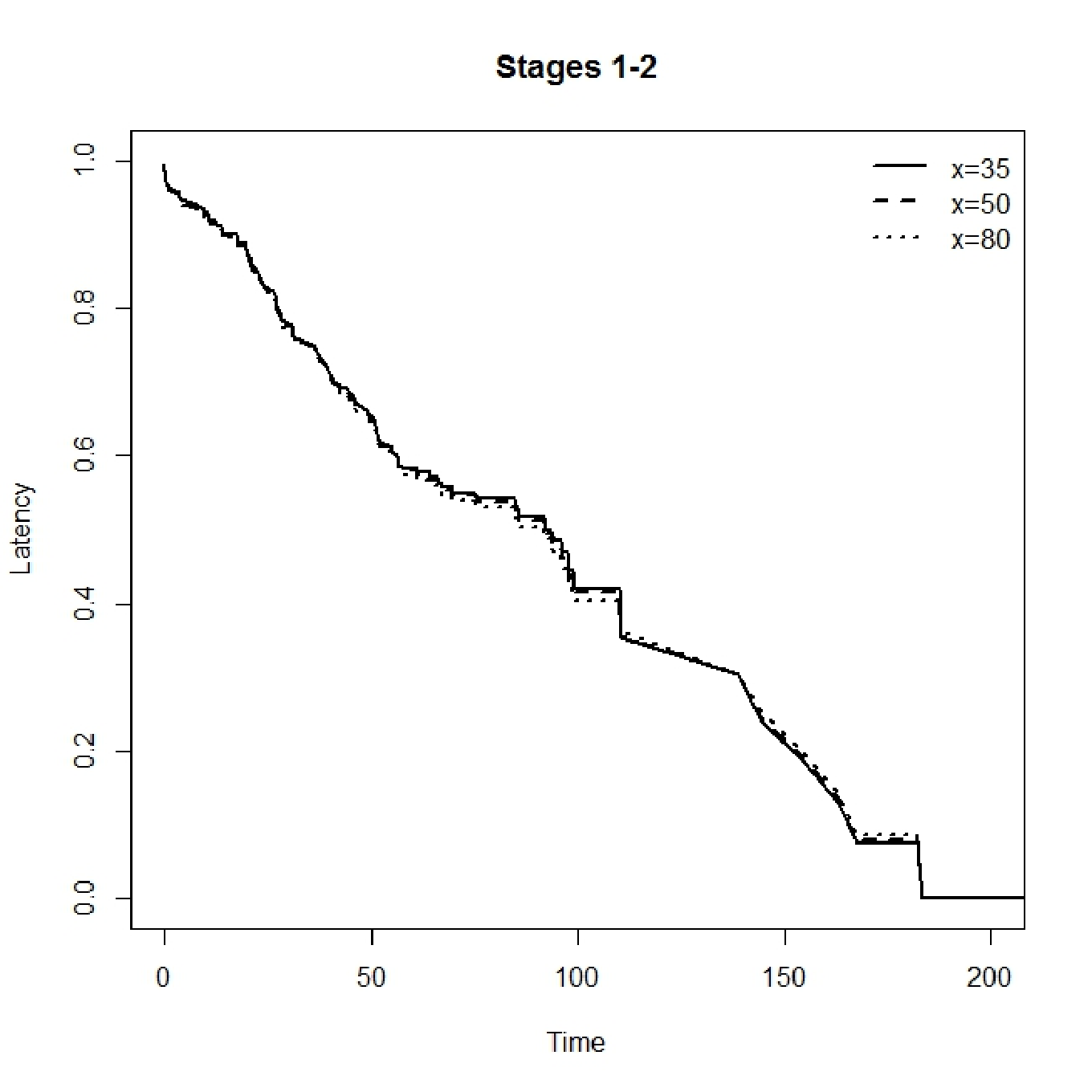}
\includegraphics[width=0.5\textwidth]{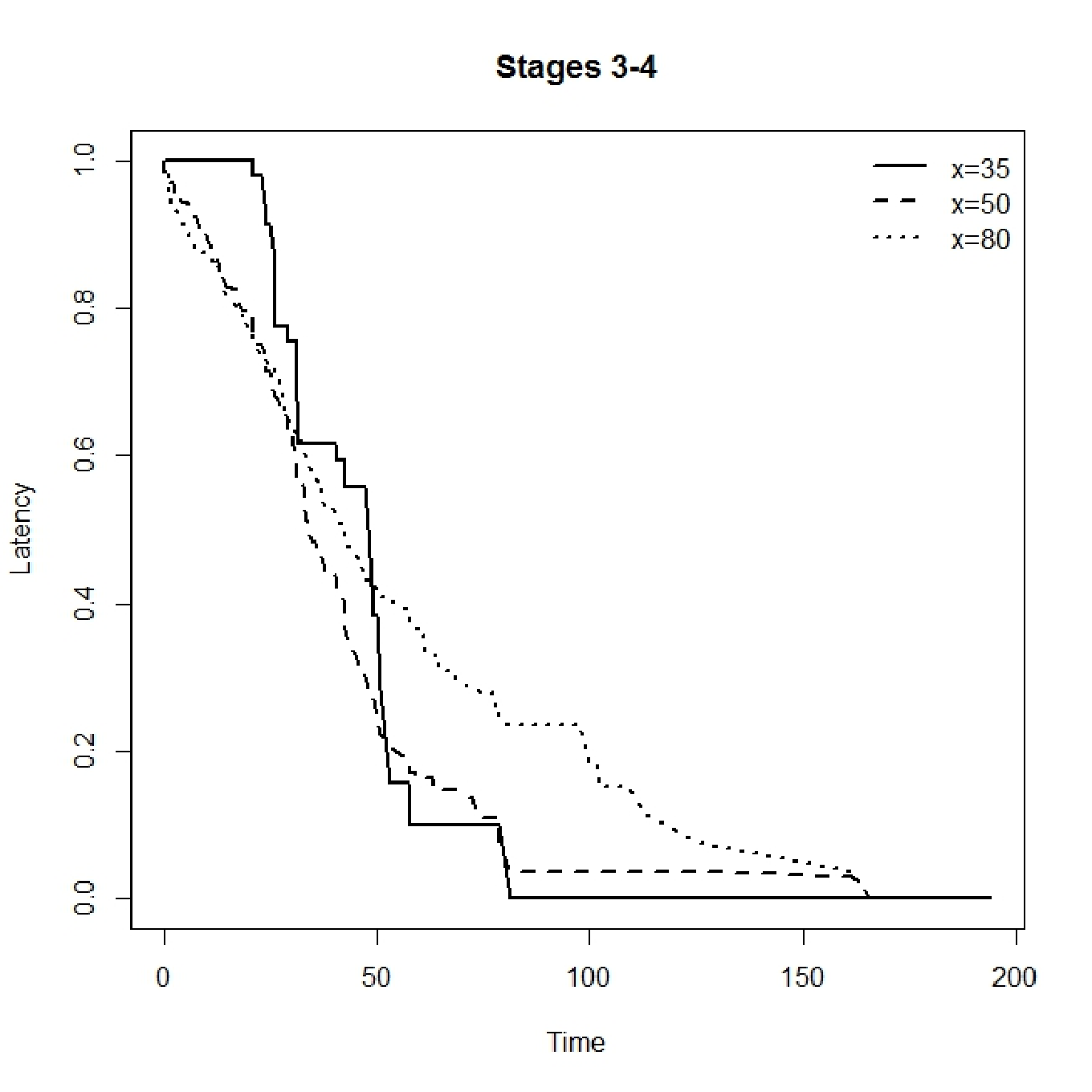}
\caption{Latency estimation for patients in Stages $1-2$ (left) and $3-4$ (right) with ages $35$ (solid line), $50$ (dashed line) and $80$ (dotted line), computed using the nonparametric estimator, $\hat S_{0,h}(t|x)$, with the bootstrap bandwidth, $h_x^*$.}
\label{fig:Latency_h_Stage}
\end{figure}

The latency estimation computed with the bootstrap bandwidth, $\hat S_{0,h^*}(t|x)$, for different ages ($35$, $50$ and $80$) is shown in Figure \ref{fig:Latency_h_Stage}. We can observe that for Stages $1-2$ the covariate age does not seem to be determining for the latency estimation, since all the estimated latency
functions are very similar for the whole grid of
ages. On the contrary, for Stages $3-4$ the latency estimation varies
considerably depending on the age. For example, the probability that
the follow-up time since the diagnostic until death is larger than $4.5$ years
($54$ months) is around $0.2$ for patients with ages $35$ and $50$, whereas
for $80$ year old patients, that probability is larger than $0.4$.

\begin{acknowledgements}
The first author's research was sponsored by the Spanish FPU (Formaci\'on de Profesorado Universitario) Grant from MECD (Ministerio de Educaci\'on, Cultura y Deporte) with reference FPU13/01371. All the authors acknowledge partial support by
the MINECO (Ministerio de Econom\'ia y Competitividad) grant MTM2014-52876-R (EU ERDF support included), the MICINN (Ministerio de Ciencia e Innovaci\'on) Grant MTM\-2011-22392 (EU ERDF support included) and Xunta de Galicia GRC Grant CN2012/\-130. The authors are grateful to Dr. Sonia P\'ertega and Dr. Salvador Pita, at the University Hospital of A Coru\~na, for providing the colorectal cancer data set.
\end{acknowledgements}

\section*{Appendix}


\noindent \textbf{Proof of Theorem \ref{thm:iid_representation}.}
The nonparametric estimator of $S_0(t|x)$ in (\ref{ec:S0_estimation}) can be
decomposed as follows:%
\begin{equation}
\hat{S}_{0,h}(t|x)-S_{0}(t|x)=A_{11}+ A_{21} +A_{12}+
A_{22},\label{S0hS0sumaA}
\end{equation}
where the dominant terms of the i.i.d. representation of $\hat
S_{0,h}(t|x)$ derive from
\begin{equation}
A_{11} =\frac{\hat{S}_{h}(t|x)-S(t|x)}{p(x)}
 \text{ and }
A_{21} =\frac{1-S(t|x)}{p^{2}(x)}(\hat{p}_{h}(x)-p(x)),
\label{ec:A21A31}
\end{equation}
and the remaining terms
\begin{equation}
A_{12} =\frac{(\hat{S}_{h}(t|x)-S(t|x))(p(x)-\hat{p}_{h}(x))}{\hat{p}_{h}(x)p(x)}
\text{ and }
A_{22}= \frac{S(t|x)-1}{p^2(x)}\frac{\left( \hat{p}%
_{h}(x)-p(x)\right) ^{2}}{\hat{p}_{h}(x)}  \label{ec:A22A32}
\end{equation}
\noindent will be proved to be negligible.

The i.i.d. representation of the term $A_{11}$ in (\ref{ec:A21A31}) follows, under
assumptions (A1)-(A7), (A11) and (A12), from that of $\hat S_h(t|x)$ in
Theorem 2 of \cite{Iglesias-Perez}:
\begin{equation}
A_{11}=-\frac{S(t|x)}{p(x)} \sum_{i=1}^{n}\tilde{B}_{h,i}(x)\xi
(T_{i},\delta _{i},t,x)+O\left( \left( \frac{\ln n}{nh}\right) ^{3/4}\right)
\text{ \ a.s.}  \label{ec:A11iid}
\end{equation}

Under assumptions (A1)-(A12), the dominant terms of the i.i.d. representation of $A_{21}$ in (\ref%
{ec:A21A31}) come from the i.i.d. representation of $\hat p_h(x)$ in Theorem 3 of \cite{LopezCheda}:
\begin{equation}
A_{21}= -\frac{(1-S(t|x))}{p^{2}(x)}(1-p(x))\sum_{i=1}^{n} \tilde{B}_{h,i}(x)\xi (T_{i},\delta _{i},\infty ,x)  
+O\left( \left( \frac{\ln n}{nh}\right) ^{3/4}\right) \text{ \ a.s.}   \label{ec:A2131iid}
\end{equation}

We continue by proving the negligibility of $A_{12}$ in (\ref{ec:A22A32}). Under
assumptions (A3a), (A4), (A5) and (A11), we apply Lemma 5 in \cite%
{Iglesias-Perez} to obtain
\begin{equation*}
\hat{S}_{h}(t|x)-S(t|x)=O\left( \sqrt{\frac{\ln \ln n}{nh}}+h^{2}\right)
\text{ a.s.}
\end{equation*}%
and, similarly from Theorem 3.3 in Arcones (1997) and the Strong Law of
Large Numbers (SLLN),
\begin{equation}
\hat{p}_{h}(x)-p(x)=O\left( \sqrt{\frac{\ln \ln n}{nh}}+h^{2}\right) \text{
a.s.}  \label{ph-p=O}
\end{equation}
It is straightforward to check that if the bandwidth satisfies $h\rightarrow
0$, $\frac{\ln n}{nh}\rightarrow 0$ and $\frac{nh^{5}}{\ln n}=O(1)$, with the convergence $\hat{p}_{h}(x)\rightarrow p(x)\text{ a.s.}$ proved in Lemma 7 of \cite{LopezCheda}, it directly follows that
\begin{equation}
A_{12}=O\left( \left( \frac{\ln n}{nh}\right) ^{3/4}\right) a.s.
\label{ec:A12iid}
\end{equation}%

With respect to $A_{22}$ in (\ref{ec:A22A32}), if $h\rightarrow 0$, $\frac{\ln n}{nh}\rightarrow 0$ and $\frac{nh^{5}}{\ln n}=O(1)$, using the almost sure consistency of $\hat{p}_{h}(x)$, it follows from (\ref{ph-p=O}) that
\begin{equation}
A_{22}=O\left( \left( \frac{\ln n}{nh}\right) ^{3/4}\right) \text{a.s.}  \label{ec:A2232iid}
\end{equation}
The proof of the theorem follows from the decomposition (\ref{S0hS0sumaA}) and the results (\ref{ec:A11iid}), (\ref{ec:A2131iid}), (\ref{ec:A12iid}) and (\ref{ec:A2232iid}).


\noindent \textbf{Proof of Theorem \ref{thm:AMSE}.}
From Theorem \ref{thm:iid_representation}, the latency estimator can be
decomposed as
\begin{equation*}
\hat{S}_{0,h}(t|x)-S_{0}(t|x)=C_{1}+C_{2}+O\left( \left( \frac{\ln n}{nh}%
\right) ^{3/4}\right) \text{ a.s.},
\end{equation*}%
\noindent where
\begin{eqnarray*}
C_{1}&=&-\frac{S(t|x)}{p(x)}\sum_{i=1}^{n}\tilde{B}_{h,i}(x)\xi (T_{i},\delta_{i},t,x),\\
C_{2}&=&-\frac{(1-p(x))(1-S(t|x))}{p^{2}(x)}\sum_{i=1}^{n}\tilde{B}%
_{h,i}(x)\xi (T_{i},\delta _{i},\infty ,x),
\end{eqnarray*}
with $\tilde{B}_{h,i}(x)$ in (\ref{ec:Btilde}) and $\xi $ in (\ref{ec:xi}).
Then, the AMSE of $\hat{S}_{0,h}(t|x)$ is
\begin{equation}
AMSE(\hat{S}_{0,h}(t|x))=E(C_{1}^{2})+E(C_{2}^{2})+2E(C_{1}\cdot C_{2}).
\label{ec:MSE_C1_C1_C1C2}
\end{equation}

We start with the first term of $AMSE(\hat{S}_{0, h} (t|x) )$. Note that
\begin{equation}  \label{ec:ec1}
E(C_{1}^{2})=Var(C_{1})+(E(C_{1}))^{2},
\end{equation}
where
\begin{equation}
Var(C_{1}) = \frac{1}{n h^2} \left( \frac{S(t|x)}{p(x)}\right) ^{2} \frac{1}{m^2(x)}
Var \left ( K \left ( \frac{x - X_1}{h} \right ) \xi(T_1, \delta_1, t, x)
\right )  \label{varC1}
\end{equation}


\noindent and
\begin{eqnarray}
&& Var \left ( K \left ( \frac{x - X_1}{h} \right ) \xi(T_1, \delta_1, t, x)
\right )   \label{ec:var_2} \\
&=& E \left ( K^2 \left ( \frac{x - X_1}{h} \right ) \xi^2(T_1, \delta_1, t,
x) \right ) - \left [ E \left ( K \left ( \frac{x - X_1}{h} \right )
\xi(T_1, \delta_1, t,x) \right ) \right ]^2.  \notag
\end{eqnarray}

Let us consider $\Phi _{1}(y,t,x)$ defined in (\ref{Phi_1}). From a change of variable and a Taylor
expansion, then the first term in (\ref{ec:var_2}) is
\begin{equation}
E\left[ K^{2}\left( \frac{x-X_{1}}{h}\right) \xi ^{2}(T_{1},\delta _{1},t,x)%
\right] =h\Phi _{1}(x,t,x)m(x)c_{K}+O(h^{3}).  \label{ec:var_2a}
\end{equation}

For the second term in (\ref{ec:var_2}), applying a change of variable, a Taylor expansion, and taking into
account the symmetry of $K$, it follows that 
\begin{equation}
\left[ E\left( K\left( \frac{x-X_{1}}{h}\right) \xi (T_{1},\delta
_{1},t,x)\right) \right] ^{2}=\left[ \Phi (x,t,x)m(x)h+O(h^{3})\right] ^{2}=O(h^{6}), \label{ec:var_2b}
\end{equation}%
\noindent where $\Phi (y,t,x)=E\left[ \xi (T,\delta ,t,x)|X=y\right] $ and, as will be proved in Lemma \ref{lem:phi}, $\Phi (x,t,x)=0$ for all $t\geq 0$.

From (\ref{varC1}), (\ref{ec:var_2}), (\ref{ec:var_2a}) and (\ref%
{ec:var_2b}), then
\begin{equation*}
Var(C_{1})=\frac{1}{nh}\left( \frac{S(t|x)}{p(x)}\right) ^{2}\frac{1}{m(x)}%
\Phi _{1}(x,t,x)c_{K}+O\left( \frac{h}{n}\right) .
\end{equation*}



Continuing with the second term in the right hand side of (\ref{ec:ec1}):
\begin{equation*}
E(C_{1}) =-\frac{1}{h}\frac{S(t|x)}{m(x)p(x)}E\left[ K\left( \frac{x-X_{1}}{h}%
\right) \xi (T_{1},\delta _{1},t,x)\right] .
\end{equation*}

Using a Taylor expansion, and $\Phi (x,t,x)=0 \; \forall t\geq 0$, then
\begin{equation*}
E(C_{1})=-\frac{1}{2}h^{2}\frac{S(t|x)}{p(x)m(x)}d_{K}\left( \Phi ^{\prime
\prime }\left( x,t,x\right) m(x)+2\Phi ^{\prime }\left( x,t,x\right)
m^{\prime }(x)\right) +o(h^{2}).
\end{equation*}

So the first term of $AMSE(\hat{S}_{0,h}(t|x))$ in (\ref{ec:MSE_C1_C1_C1C2})
is
\begin{eqnarray}
E(C_{1}^{2}) &=&
\frac{1}{4}h^{4}d_{K}^{2}\left[ \frac{S(t|x)}{p(x)m(x)}\left( \Phi
^{\prime \prime }\left( x,t,x\right) m(x)+2\Phi ^{\prime }\left(
x,t,x\right) m^{\prime }(x)\right) \right] ^{2} \notag \\
&&+\frac{1}{nh}\left( \frac{S(t|x)}{p(x)}\right) ^{2}\frac{1}{%
m(x)}\Phi _{1}(x,t,x)c_{K}  +o(h^{4})+O\left( \frac{h}{n}%
\right).  \label{EC12}
\end{eqnarray}


Following the same ideas as those for $C_{1}$, we obtain for $C_{2}$ that
\begin{eqnarray}
E(C_{2}^{2}) &=&\frac{1}{nh}\left( \frac{(1-S(t|x))(1-p(x))}{p^{2}(x)}%
\right) ^{2}\frac{1}{m(x)}\Phi _{1}(x,\infty ,x)c_{K}  \notag \\
&+&\frac{1}{4}h^{4}d_{K}^{2}\left[ \frac{(1-S(t|x))(1-p(x))}{p^{2}(x)m\left(
x\right) } \right. \notag \\
&\times&  \left. \left(  \Phi ^{\prime \prime }\left( x,\infty ,x\right) m(x)+2\Phi
^{\prime }\left( x,\infty ,x\right) m^{\prime }(x)\right) \right] ^{2}o(h^{4})+O\left( \frac{h}{n}\right) .
\label{EC22} 
\end{eqnarray}


We continue studying the third term of $AMSE(\hat{S}_{0,h}(t|x))$ in (\ref%
{ec:MSE_C1_C1_C1C2}):
\begin{equation*}
E\left( C_{1}\cdot C_{2}\right)=\frac{(1-p(x))S(t|x)(1-S(t|x)}{p^{3}(x)}\left[ n(n-1)\alpha \beta+n\gamma \right] ,
\end{equation*}
where
\begin{eqnarray*}
\alpha &=&E\left[ \tilde{B}_{h1}(x)\xi (T_{1},\delta _{1},t,x)\right] , \\
\beta &=&E\left[ \tilde{B}_{h1}(x)\xi (T_{1},\delta _{1},\infty ,x)\right]
\text{,} \\
\gamma &=&E\left[ \tilde{B}_{h1}^{2}(x)\xi (T_{1},\delta _{1},t,x)\xi
(T_{1},\delta _{1},\infty ,x)\right] .
\end{eqnarray*}%
\noindent  Using a Taylor expansion and $\Phi (x,t,x)=0$ for all $t\geq 0$, the terms $\alpha $ and $\beta $ are
\begin{eqnarray}
\alpha &=&\frac{1}{2}\frac{h^{2}}{n}d_{K}\frac{1}{m(x)}\left( \Phi ^{\prime
\prime }\left( x,t,x\right) m(x)+2\Phi ^{\prime }\left( x,t,x\right)
m^{\prime }(x)\right) +o\left( \frac{h^{2}}{n}\right),  \label{alfa} \\
\beta &=&\frac{1}{2}\frac{h^{2}}{n}d_{K}\frac{1}{m(x)}\left( \Phi ^{\prime
\prime }\left( x,\infty ,x\right) m(x)+2\Phi ^{\prime }\left( x,\infty
,x\right) m^{\prime }(x)\right) +o\left( \frac{h^{2}}{n}\right).
\label{beta}
\end{eqnarray}%
For the term $\gamma $, it follows that
\begin{eqnarray}
\gamma &=&\frac{1}{n^{2}h^{2}}\frac{1}{m^{2}(x)}\int K^{2}\left( \frac{x-y}{h}%
\right) \Phi _{2}(y,t,x)m(y)dy, \notag\\
&=&\frac{1}{n^{2}h}\frac{1}{m(x)}\Phi _{2}(x,t,x)c_{K}+O\left( \frac{h}{%
n^{2}}\right),\label{gamma}
\end{eqnarray}
where $\Phi _{2}(y,t,x)=E\left[ \xi (T,\delta ,t,x)\xi (T,\delta ,\infty ,x)|X=y\right] .$
From (\ref{alfa}), (\ref{beta}) and (\ref{gamma}), the third term of $AMSE(%
\hat{S}_{0,h}(t|x))$ in (\ref{ec:MSE_C1_C1_C1C2}) is:
\begin{eqnarray}
&&E\left( C_{1}\cdot C_{2}\right) = \frac{(1-p(x))S(t|x)(1-S(t|x)}{p^{3}(x)}\left[\frac{1}{4}h^{4}d_{K}^{2}\frac{1}{m^{2}(x)}\right. \notag\\
&&\times\left( \Phi ^{\prime \prime
}\left( x,t,x\right) m(x)+2\Phi ^{\prime }\left( x,t,x\right) m^{\prime
}(x)\right) \left( \Phi ^{\prime \prime }\left( x,\infty ,x\right)
m(x)+2\Phi ^{\prime }\left( x,\infty ,x\right) m^{\prime }(x)\right)   \notag
\\
&&+\left.\frac{1}{nh}\frac{1}{m(x)}\Phi _{2}(x,t,x)c_{K} \right] +o\left( h^{4}\right) +O\left( \frac{h}{n}\right).
\label{E_C1C2}
\end{eqnarray}

Compiling (\ref{EC12}), (\ref{EC22}) and (\ref{E_C1C2}), the $AMSE(\hat{S}
_{0,h}(t|x))$ in (\ref{ec:MSE_C1_C1_C1C2}) is 
\begin{eqnarray*}
&&AMSE(\hat{S}_{0,h}(t|x))\\
&=&\frac{1}{nh}\frac{1}{m(x)}c_{K}
\left( \left( \frac{S(t|x)}{p(x)}\right)^{2}\Phi _{1}(x,t,x)
+ \left( \frac{(1-S(t|x))(1-p(x))}{p^{2}(x)}\right)^{2}\Phi _{1}(x,\infty ,x)\right.\\
&&+\left. 2\frac{(1-p(x))S(t|x)(1-S(t|x))}{p^{3}(x)}\Phi _{2}(x,t,x)\right)  \\
&+&\frac{1}{4}h^{4}d_{K}^{2}\frac{1}{m^{2}(x)}\left( \frac{S(t|x)}{p(x)}\left( \Phi ^{\prime \prime }\left( x,t,x\right) m(x)+ 2 \Phi ^{\prime }\left( x,t,x\right) m^{\prime }(x)\right)\right.\\
&+&\left. \frac{(1-S(t|x))(1-p(x))}{p^{2}(x)}  \left( \Phi ^{\prime \prime
}\left( x,\infty ,x\right) m(x)+2\Phi ^{\prime }\left( x,\infty ,x\right)
m^{\prime }(x)\right) \right) ^{2}  + o(h^{4})+O\left( \frac{h}{n}\right). 
\end{eqnarray*}

Since, from (\ref{Phi1(x,t,x)}) and (\ref{Phi2(x,t,x)}), in Lemmas \ref{lem:phi1} and \ref{lem:phi2} it is proven that
\begin{equation*}
\Phi _{1}(x,t,x)=\Phi _{2}(x,t,x)=\int_{0}^{t}\frac{dH^{1}\left( v|x\right)
}{\left( 1-H(v|x)\right) ^{2}},
\end{equation*}%
and considering (\ref{ec:B1})-(\ref{ec:V3}),
the AMSE of $\hat{S}_{0,h}(t|x)$ is, finally, that in (\ref{th2:amse}). 

This completes the proof.

\begin{lem}
\label{lem:phi}
The term $\Phi \left(y,t,x\right)$ in (\ref{Phi}) has the following expression:
\begin{equation*}
\Phi \left( y,t,x\right) =\int_{0}^{t}\frac{dH^{1}\left( v|y\right) }{%
1-H(v|x)}-\int_{0}^{t}(1-H(v|y))\frac{dH^{1}(v|x)}{\left( 1-H(v|x)\right)
^{2}},
\end{equation*}
and consequently, $\Phi \left(x,t,x\right) =0$ for any $t\geq 0$.

\end{lem}

\noindent \textbf{Proof of Lemma \ref{lem:phi}.}
Let us recall $\Phi \left( y,t,x\right) =E\left[ \xi (T,\delta ,t,x)|X=y\right] $, then
\begin{eqnarray*}
\Phi \left( y,t,x\right) &=& E\left[ \frac{1\{T\leq t,\delta =1\}}{1-H(T|x)}\bigg
|X=y\right] -E\left[ \int_{0}^{t}\frac{\left( y\leq T\right) dH^{1}(u|x)}{%
\left( 1-H(u|x)\right) ^{2}}\bigg|X=y\right] \\
&=& A^{\prime }-A^{\prime \prime }.
\end{eqnarray*}%
We start with $A^{\prime }$:
\begin{equation*}
A^{\prime } =E\left[ \frac{1\{T\leq t\}}{1-H(T|x)}E\left( \delta |T,X=y\right) \right] 
=\int\limits_{0}^{t}\frac{q(v,y)dH(v|y)}{1-H(v|x)}=\int_{0}^{t}\frac{%
dH^{1}\left( v|y\right) }{1-H(v|x)},
\end{equation*}
where $q\left( t,y\right) =E\left( \delta |T=t,X=y\right)$ and $H_{1}\left( t|y\right) =P\left( T\leq t,\delta =1|X=y\right)$. \\
We continue with $A^{\prime \prime }$:
\begin{equation*}
A^{\prime \prime }=\int_{0}^{t}E\left[ 1\{v\leq
T\}|X=y\right] \frac{dH^{1}(v|x)}{\left( 1-H(v|x)\right) ^{2}} =\int_{0}^{t}(1-H(v|y))\frac{dH^{1}(v|x)}{\left( 1-H(v|x)\right) ^{2}}.
\end{equation*}

Then,
\begin{equation}
\Phi \left( y,t,x\right) =\int_{0}^{t}\frac{dH^{1}\left( v|y\right) }{%
1-H(v|x)}-\int_{0}^{t}(1-H(v|y))\frac{dH^{1}(v|x)}{\left( 1-H(v|x)\right)
^{2}},  \label{Phi(y,t,x)}
\end{equation}%
and therefore, $\Phi \left( x,t,x\right) =0$ for any $t\geq 0$.

\begin{lem}
\label{lem:phi1}
The term $\Phi_1(y,t,x)$ in (\ref{Phi_1}) verifies, for any $t \in [a,b]$,
\begin{equation}
\Phi _{1}\left( x,t,x\right) =\int_{0}^{t}\frac{dH^{1}\left( v|x\right) }{\left( 1-H(v|x)\right) ^{2}}.\label{Phi1(x,t,x)}
\end{equation}

\end{lem}

\noindent \textbf{Proof of Lemma \ref{lem:phi1}.}
Note that $\Phi_{1}\left( y,t,x\right) =E\left[ \xi ^{2}(T,\delta ,t,x)|X=y\right]$, with $\xi$ in (\ref{ec:xi}). Then,
\begin{eqnarray*}
\Phi _{1}\left( y,t,x\right) &=&E\left[ \frac{1\{T\leq t,\delta =1\}}{\left(
1-H(T|x)\right) ^{2}}\bigg |X=y\right] \\
&+&E\left[ \int_{0}^{t}\int_{0}^{t}\frac{1 \{ u\leq T \} 1\{ v\leq
T \} }{\left( 1-H(u|x)\right) ^{2}\left( 1-H(v|x)\right) ^{2}}dH^{1}(u|x)dH^{1}(v|x)\bigg |X=y\right] \\
&-&2E\left[ \frac{1\{T\leq t,\delta =1\}}{1-H(T|x)}\int_{0}^{t}\frac{1\{u\leq T \} dH^{1}(u|x)}{\left( 1-H(u|x)\right) ^{2}}\bigg |X=y\right]
\\
&=&A+B-2C.
\end{eqnarray*}%
The first term in the decomposition of $\Phi _{1}\left( y,t,x\right) $ is%
\begin{equation*}
A=\int_{0}^{t}\frac{q\left( v,y\right) }{\left( 1-H(v|x)\right) ^{2}}%
dH(v|y)=\int_{0}^{t}\frac{dH^{1}\left( v|y\right) }{\left( 1-H(v|x)\right)
^{2}}.
\end{equation*}

The second term is
\begin{equation*}
B=\int_{0}^{t}\int_{0}^{t}\frac{1-H\left( \max \left( w,v\right) |y\right)
}{\left( 1-H(v|x)\right) ^{2}\left( 1-H(w|x)\right) ^{2}}%
dH^{1}(v|x)dH^{1}(w|x).
\end{equation*}

Integrating in the supports $\left\{ (v,w) \in \left[ 0,t\right]  \times \left[ 0,t\right]    /v\leq
w\right\} $ and $\left\{ \left ( v,w \right ) \in  \right.$ \\ $\left.\left[ 0,t\right] \times   \left[ 0,t\right]  /w < v\right\} $, the
term $B$ is%
\begin{equation*}
B=2\int_{0}^{t}\frac{1}{\left( 1-H(v|x)\right) ^{2}}\left( \int_{v}^{t}%
\frac{1-H\left( w|y\right) }{\left( 1-H(w|x)\right) ^{2}}dH^{1}(w|x)\right)
dH^{1}(v|x).
\end{equation*}

Finally, the third term in the decomposition of $\Phi _{1}\left(y,t,x\right) $ is
\begin{equation*}
C=\int_{0}^{t}\frac{1}{\left( 1-H(u|x)\right) ^{2}}\left( \int_{u}^{t}\frac{%
dH^{1}\left( v|y\right) }{1-H(v|x)}\right) dH^{1}(u|x).
\end{equation*}

Note that, for $y=x$, we have that $B=2C$. This completes the proof.

\begin{lem}
\label{lem:phi2}

The expression for the term $\Phi_{2}(x,t,x)$, for any $t \in [a,b]$, is the following:
\begin{equation}
\label{Phi2(x,t,x)}
\Phi_{2}(x,t,x)=\int_{0}^{t}\frac{dH^{1}\left( v|x\right) }{\left(1-H(v|x)\right) ^{2}}.
\end{equation}

\end{lem}

\noindent \textbf{Proof of Lemma \ref{lem:phi2}.}
Recall $\Phi _{2}(y,t,x)=E\left[ \xi \left( T,\delta ,t,x\right) \xi(T,\delta ,\infty ,x)|X=y\right]$ with $\xi$ in (\ref{ec:xi}). Then:
\begin{eqnarray*}
&&\Phi _{2}(y,t,x) \\
&=&E\left[ \frac{1\{T\leq t,\delta =1\}}{\left(
1-H(T|x)\right) ^{2}}\bigg |X=y\right] -E\left[ \frac{1\{\delta =1\}}{1-H(T|x)}\int_{0}^{\infty }\frac{1\{u\leq T\leq t\} }{\left( 1-H(u|x)\right) ^{2}}dH^{1}(u|x)\bigg |X=y
\right] \\
&-&E\left[ \frac{1\{\delta =1\}}{1-H(T|x)}\int_{0}^{t}\frac{1\{ v\leq T\} }{\left( 1-H(v|x)\right) ^{2}}dH^{1}(v|x)\bigg |X=y\right] \\
&+&E\left[ \int_{0}^{t}\frac{1\{ v\leq T \} dH^{1}(v|x)}{\left(
1-H(v|x)\right) ^{2}}\int_{0}^{\infty }\frac{1\{ u\leq T\}
dH^{1}(u|x)}{\left( 1-H(u|x)\right) ^{2}}\bigg |X=y\right] \\
&=&A-B-C+D.
\end{eqnarray*}
Straightforward calculations yield:
\begin{eqnarray*}
A&=&\int_{0}^{t}\frac{dH^{1}\left( v|y\right) }{\left(
1-H(v|x)\right) ^{2}},\\
B&=&\int_{0}^{\infty }\left( \int_{u}^{t}\frac{dH^{1}\left( v|y\right) }{%
1-H(v|x)}\right) \frac{dH^{1}(u|x)}{\left( 1-H(u|x)\right) ^{2}},\\
C&=&\int_{0}^{t}\left( \int_{v}^{\infty }\frac{dH^{1}\left( u|y\right) }{%
1-H(u|x)}\right) \frac{dH^{1}(v|x)}{\left( 1-H(v|x)\right) ^{2}},\\
D&=&\int_{0}^{t}\frac{1}{\left( 1-H(v|x)\right) ^{2}}\left( \int_{0}^{\infty }%
\frac{1-H\left( \max \left( u,v\right) |y\right) }{\left( 1-H(u|x)\right)
^{2}}dH^{1}(u|x)\right) dH^{1}(v|x).
\end{eqnarray*}

Integrating in the supports $\left\{\left ( u,v \right )\in \left[ 0,\infty \right) \times \left [0,t \right ] /v\leq u\right\}
$ and $\left\{ \left ( u,v \right )  \in  \right .$\\
$ \left. \left[ 0,\infty \right )   \times \left [0,t \right] /u < v\right\} =\left\{ \left ( u, v \right ) \in \left[ 0, t \right ] \times  \left [0, t \right ] /u < v \right\} $, the term $D$ is%
\begin{eqnarray*}
D &=&\int_{0}^{t}\left( \int_{v}^{\infty }\frac{1-H\left( u|y\right) }{%
\left( 1-H(u|x)\right) ^{2}}dH^{1}(u|x)\right) \frac{dH^{1}(v|x)}{\left(
1-H(v|x)\right) ^{2}} \\
&+&\int_{0}^{\infty }\left( \int_{u}^{t}\frac{1-H\left( v|y\right) }{\left(
1-H(v|x)\right) ^{2}}dH^{1}(v|x)\right) \frac{dH^{1}(u|x)}{\left(
1-H(u|x)\right) ^{2}}.
\end{eqnarray*}%
When $y=x$, then $D=C+B$, which concludes the proof.

\noindent \textbf{Proof of Theorem \ref{thm:asymptotic_normality}.}
Under assumptions (A1)-(A10) and using Theorem \ref{thm:iid_representation}, $\sqrt{nh}\left( \hat{S}_{0,h}(t|x)-S_{0}(t|x)\right) $ has the same limit distribution as
\begin{equation*}
\sqrt{nh}\sum_{i=1}^{n}\eta _{h}(T_{i},\delta _{i},X_{i},t,x)=-\left( I+II+III+IV\right) ,
\end{equation*}
where
\begin{eqnarray*}
I &=&\sqrt{nh}\frac{1}{nh}\frac{S(t|x)}{p(x)m(x)} \\
&\times & \sum_{i=1}^{n}\left[
K\left( \frac{x-X_{i}}{h}\right) \xi (T_{i},\delta _{i},t,x)-E\left( K\left(
\frac{x-X_{i}}{h}\right) \xi (T_{i},\delta _{i},t,x)\right) \right] , \\
II &=&\sqrt{nh}\frac{1}{nh}\frac{(1-p(x))(1-S(t|x))}{p^{2}(x)m\left(
x\right) } \\
&\times & \sum_{i=1}^{n}\left[ K\left( \frac{x-X_{i}}{h}\right) \xi
(T_{i},\delta _{i},\infty ,x)-E\left( K\left( \frac{x-X_{i}}{h}\right) \xi
(T_{i},\delta _{i},\infty ,x)\right) \right] , \\
III &=&\sqrt{nh}\frac{1}{nh}\frac{S(t|x)}{p(x)m(x)}\sum_{i=1}^{n}E\left[
K\left( \frac{x-X_{i}}{h}\right) \xi (T_{i},\delta _{i},t,x)\right] , \\
IV &=&\sqrt{nh}\frac{1}{nh}\frac{(1-p(x))(1-S(t|x))}{p^{2}(x)m\left(
x\right) }\sum_{i=1}^{n}E\left[ K\left( \frac{x-X_{i}}{h}\right) \xi
(T_{i},\delta _{i},\infty ,x)\right] .
\end{eqnarray*}

The deterministic part $b(t,x)$ comes from $III+IV$. Recall the function $\Phi (y,t,x)$ in (\ref{Phi(y,t,x)}), since $\Phi (x,t,x)=0$, then
\begin{eqnarray}
&& E\left[ K\left( \frac{x-X}{h}\right) \xi (T,\delta ,t,x)\right ]  \notag \\
&=& \frac{1}{2} h^{3}d_{K}\left( \Phi ^{\prime \prime }(x,t,x)m(x) + 2\Phi ^{\prime }(x,t,x)m^{\prime }(x)\right) +o(h^{3}). \label{EKpsi}
\end{eqnarray}
Therefore,
\begin{eqnarray*}
III&=&\sqrt{nh^{5}}\frac{S(t|x)}{p(x)m(x)}\frac{1}{2}d_{K}\left( \Phi ^{\prime
\prime }(x,t,x)m(x)+2\Phi ^{\prime }(x,t,x)m^{\prime }(x)\right) \left(
1+o\left( 1\right) \right),\\
IV&=&\sqrt{nh^{5}}\frac{(1-p(x))(1-S(t|x))}{p^{2}(x)m\left( x\right) }\frac{%
1}{2}d_{K}\left( \Phi ^{\prime \prime }(x,\infty ,x)m(x) + 2\Phi ^{\prime}(x,\infty ,x)m^{\prime }(x)\right) \left( 1+o(1)\right) .
\end{eqnarray*}

If  $nh^{5}\rightarrow 0$, then $III+IV=o\left( 1\right) $ and $b\left(
t,x\right) =0$. On the other hand, if $nh^{5}\rightarrow C^{5}$ then
\begin{eqnarray*}
b(t,x) &=&C^{5/2}\frac{S(t|x)}{p(x)m(x)}\frac{1}{2}d_{K}\left( \Phi ^{\prime
\prime }(x,t,x)m(x)+2\Phi ^{\prime }(x,t,x)m^{\prime }(x)\right) \\
&+&C^{5/2}\frac{(1-p(x))(1-S(t|x))}{p^{2}(x)m\left( x\right) }\frac{1}{2}%
d_{K}\left( \Phi ^{\prime \prime }(x,\infty ,x)m(x)+ 2\Phi ^{\prime }(x,\infty ,x)m^{\prime }(x)\right).
\end{eqnarray*}%
As for the asymptotic distribution of $I+II$, it is immediate to prove that:
\begin{equation*}
I+II=\sum_{i=1}^{n}\left( \gamma _{i,n}(x,t)+\Gamma _{i,n}(x,t)\right) ,
\end{equation*}%
where%
\begin{eqnarray*}
\gamma _{i,n}(x,t) &=&\frac{1}{\sqrt{nh}}\frac{S(t|x)}{p(x)m(x)}\\
&\times& \left[ K\left( \frac{x-X_{i}}{h}\right) \xi (T_{i},\delta
_{i},t ,x)-E\left( K\left( \frac{x-X_{i}}{h}\right) \xi (T_{i},\delta
_{i},t ,x)\right) \right], \\
\Gamma _{i,n}(x,t) &=&\frac{1}{\sqrt{nh}}\frac{(1-p(x))(1-S(t|x))}{%
p^{2}(x)m\left( x\right) } \\
&\times& \left[ K\left( \frac{x-X_{i}}{h}\right) \xi (T_{i},\delta
_{i},\infty ,x)-E\left( K\left( \frac{x-X_{i}}{h}\right) \xi (T_{i},\delta
_{i},\infty ,x)\right) \right],
\end{eqnarray*}%
are $n$ independent variables with mean $0$. To prove the asymptotic
normality of $I+II$, it is only necessary to show that $\sigma
_{i,n}^{2}\left( x,t\right) =Var\left( \gamma _{i,n}(x,t)+\Gamma
_{i,n}(x,t)\right)$\\ $ <\infty $, $\sigma _{n}^{2}\left( x,t\right)
=\sum_{i=1}^{n}\sigma _{i,n}^{2}\left( x,t\right) $ is positive and that the
Lindeberg condition is sa\-tis\-fied, so Lindeberg's theorem for triangular
arrays (Theorem 7.2 in \cite{Billingsley1968}, p. 42) can be applied to obtain%
\begin{equation*}
\frac{\sum_{i=1}^{n}\left( \gamma _{i,n}(x,t)+\Gamma _{i,n}(x,t)\right) }{%
\sigma _{n}\left( x,t\right) }\rightarrow N\left( 0,1\right),
\end{equation*}%
and consequently,%
\begin{equation*}
\frac{\sqrt{nh}\sum_{i=1}^{n}\eta _{h}(T_{i},\delta _{i},X_{i},t,x)}{\sigma
_{n}\left( x,t\right) }\rightarrow N\left( 0,1\right) .
\end{equation*}

We will start proving that the variance
\begin{equation}
\sigma _{i,n}^{2}\left( x,t\right) =Var\left( \gamma _{i,n}(x,t)\right)
+Var\left( \Gamma _{i,n}(x,t)\right) +2Cov\left( \gamma _{i,n}(x,t),\Gamma
_{i,n}(x,t)\right)   \label{sigma2_in}
\end{equation}%
is finite. Note that%
\begin{eqnarray*}
Var\left( \gamma _{i,n}(x,t)\right)&=&\frac{1}{nh}\left( \frac{S(t|x)}{p(x)m(x)}\right) ^{2}\left\{ E\left[
K^{2}\left( \frac{x-X_{1}}{h}\right) \xi ^{2}(T_{1},\delta _{1},t,x)\right]
\right.  \\
&&\left. -E\left[ K\left( \frac{x-X_{1}}{h}\right) \xi (T_{1},\delta
_{1},t,x)\right] ^{2}\right\} .
\end{eqnarray*}%
Let us define $\Phi _{1}(y,t,x)=E\left[ \xi ^{2}(T,\delta ,t,x)|X=y\right] $, using (\ref{EKpsi}), then the first term in (\ref{sigma2_in}) is
\begin{equation}
Var\left( \gamma _{i,n}(x,t)\right) =\frac{1}{n}\left( \frac{S(t|x)}{p(x)}%
\right) ^{2}\frac{\Phi _{1}(x,t,x)}{m\left( x\right) }c_{K}+O\left( \frac{%
h^{2}}{n}\right) .  \label{var_gi}
\end{equation}

In a similar way, the second term in (\ref{sigma2_in}) is%
\begin{equation}
Var\left( \Gamma _{i,n}(x,t)\right)=\frac{1}{n}\left( \frac{(1-p(x))(1-S(t|x))}{p^{2}(x)}\right) ^{2}\frac{%
\Phi _{1}(x,\infty ,x)}{m\left( x\right) }c_{K}+O\left( \frac{h^{2}}{n}%
\right) .  \label{var_Gi}
\end{equation}

Finally, for the third term in (\ref{sigma2_in}),%
\begin{eqnarray*}
&&Cov\left( \gamma _{i,n}(x,t),\Gamma _{i,n}(x,t)\right)  \\
&=&\frac{1}{nh}\left\{ E\left[ K^{2}\left( \frac{x-X_{i}}{h}\right) \xi
(T_{i},\delta _{i},\infty ,x)\xi (T_{i},\delta _{i},t,x)\right] \right.  \\
&&\left. -E\left[ K\left( \frac{x-X_{i}}{h}\right) \xi (T_{i},\delta
_{i},t,x)\right] E\left[ K\left( \frac{x-X_{i}}{h}\right) \xi (T_{i},\delta
_{i},\infty ,x)\right] \right\} .
\end{eqnarray*}
Let us consider $\Phi _{2}(y,t,x)=E\left[ \xi (T,\delta ,t,x)\xi (T,\delta
,\infty ,x)|X=y\right] $. Applying Taylor expansions, the third term in (\ref{sigma2_in}) is%
\begin{equation}
Cov\left( \gamma _{i,n}(x,t),\Gamma _{i,n}(x,t)\right) =\frac{1}{n}\frac{%
(1-p(x))S(t|x)(1-S(t|x))}{p^{3}(x)m\left( x\right) }\Phi
_{2}(x,t,x)c_{K}+O\left( \frac{h}{n}\right) .  \label{cov_giGi}
\end{equation}

The results (\ref{var_gi}), (\ref{var_Gi}) and (\ref{cov_giGi}), together
with (\ref{Phi1(x,t,x)}) and (\ref{Phi2(x,t,x)}), lead to
\begin{equation*}
\sigma _{i,n}^{2}\left( x,t\right) =\frac{c_{K}}{n}\left( V_{1}\left(
t,x\right) +V_{2}\left( t,x\right) +2V_{3}\left( t,x\right) \right) +O\left(
\frac{h}{n}\right),
\end{equation*}%
where $V_{1}\left( t,x\right)$, $V_{2}\left( t,x\right)$ and $V_{3}\left( t,x\right)$ are defined in (\ref{ec:V1}), (\ref{ec:V2}) and (\ref{ec:V3}), res\-pec\-ti\-vely. As a consequence, $\sigma _{i,n}^{2}\left( x,t\right)
<\infty $. The finiteness of the variance $\sigma _{n}^{2}\left( x,t\right) $
is also proved, since
\begin{equation*}
\sigma _{n}^{2}\left( x,t\right) =\sum_{i=1}^{n}\sigma _{i,n}^{2}\left(
x,t\right) =V_{1}\left( t,x\right) c_{K}+V_{2}\left( t,x\right)
c_{K}+2V_{3}\left( t,x\right) c_{K}+O\left( h\right) <+\infty .
\end{equation*}

We continue studying Lindeberg's condition:
\begin{equation}
\frac{1}{\sigma _{n}^{2}\left( x,t\right) }\sum_{i=1}^{n}\int_{\{|\gamma
_{i,n}(x,t)+\Gamma _{i,n}(x,t)|>\epsilon \sigma _{n}\left( x,t\right)
\}}(\gamma _{i,n}(x,t)+\Gamma _{i,n}(x,t))^{2}dP\rightarrow 0,\forall
\epsilon >0.  \label{Lind_cond}
\end{equation}

\noindent Let us define the indicator function $I_{i,n}\left( x,t\right)1 \left \{\left( \gamma_{i,n}(x,t)+\Gamma _{i,n}(x,t)\right) ^{2}>  \epsilon ^{2}\sigma_{n}^{2}\left( x,t\right) \right \}$.
Then (\ref{Lind_cond}) can be expressed as
\begin{equation*}
\frac{1}{\sigma _{n}^{2}\left( x,t\right) }E\left[ \sum_{i=1}^{n}(\gamma
_{i,n}(x,t)+\Gamma _{i,n}(x,t))^{2}I_{i,n}\left( x,t\right) \right] =\frac{1%
}{\sigma _{n}^{2}\left( x,t\right) }E\left( \eta _{n}\left( x,t\right)
\right),
\end{equation*}%
with
\begin{equation*}
\eta _{n}\left( x,t\right) =\sum_{i=1}^{n}(\gamma _{i,n}(x,t)+\Gamma
_{i,n}(x,t))^{2}I_{i,n}\left( x,t\right) .
\end{equation*}

Since $\frac{1}{nh}\rightarrow 0$, and the functions $K$ and $\xi $ are
bounded, one has:
\begin{eqnarray*}
&&\exists n_{0}\in \mathbb{N}/n\geq n_{0}\Rightarrow I_{i,n}(w)=0,\forall w%
\text{ and }\forall i\in \{1,2,\dots ,n\} \\
&\Leftrightarrow &\exists n_{0}\in \mathbb{N}/n\geq n_{0}\Rightarrow \eta
_{n}(w)=0,\forall w .
\end{eqnarray*}

Since $\eta _{n}(x,t)$ is bounded, then the previous condition implies that 
$ \exists n_{0}\in \mathbb{N}/n\geq n_{0}\Rightarrow E(\eta _{n}(x,t))=0$,
and then $\lim_{n\rightarrow \infty }\frac{1}{\sigma _{n}^{2}}E(\eta _{n}(x,t))=0.$
Therefore, Lindeberg's condition is proved. All these previous arguments lead to the proof of Theorem \ref{thm:asymptotic_normality}.

\bibliographystyle{spbasic}      
\bibliography{mybibfile}   


\end{document}